\documentclass[%
superscriptaddress,
 amsmath,amssymb,
 aps,twocolumn,
 pra,
]{revtex4-2}

\usepackage[normalem]{ulem}

\usepackage{graphicx}
\usepackage{dcolumn}
\usepackage{bm}
\usepackage{xcolor}
\usepackage{physics}
\usepackage{subcaption}
\usepackage{times,txfonts}
\usepackage{upgreek}
\usepackage{hyperref}

\renewcommand{\a}{\hat{a}}
\newcommand{\ad}{\hat{a}^\dagger}

\newcommand{\idd}{\mathbb{I}}

\begin{document}


\title{Thermodynamics of a phaseonium-driven optomechanical Otto engine}

\author{Federico Amato}
\email{federico.amato01@unipa.it}
\affiliation{Dipartimento di Ingegneria, Università degli Studi di Palermo, Viale delle Scienze, 90128 Palermo, Italy}
\affiliation{
    School of Mathematical Sciences and Centre for the Mathematics and Theoretical Physics of Quantum Non-Equilibrium Systems, University of Nottingham, University Park, Nottingham NG7 2RD, UK
}

\author{Gerardo Adesso}
\email{gerardo.adesso@nottingham.ac.uk}
\affiliation{
    School of Mathematical Sciences and Centre for the Mathematics and Theoretical Physics of Quantum Non-Equilibrium Systems, University of Nottingham, University Park, Nottingham NG7 2RD, UK
}

\author{G. Massimo Palma}%
\affiliation{%
    Dipartimento di Fisica e Chimica - Emilio Segrè, Università degli Studi di Palermo, Via Archirafi 36, I-90123 Palermo, Italy
}

\author{Salvatore Lorenzo}%
\affiliation{%
    Dipartimento di Fisica e Chimica - Emilio Segrè, Università degli Studi di Palermo, Via Archirafi 36, I-90123 Palermo, Italy
}

\author{Rosario Lo Franco}
\email{rosario.lofranco@unipa.it}
\affiliation{Dipartimento di Ingegneria, Università degli Studi di Palermo, Viale delle Scienze, 90128 Palermo, Italy}%


\date{\today}

\begin{abstract}
    We study an optomechanical Otto engine whose working medium is a single-mode cavity driven by beams of coherently prepared three-level phaseonium atoms. The atoms are not thermal reservoirs in the Gibbs sense; rather, their populations and ground-state coherence set the detailed-balance ratio of the cavity collision map, so that the field relaxes to a Gibbs state at an operational apparent temperature. We combine the finite-time collision-model dynamics with radiation-pressure work extraction and compare three reservoir preparations: a thermal reference at the same apparent temperatures, an incoherent atomic beam with the same populations, and the coherent phaseonium beam. We show that the phaseonium isochore charges the cavity passively: the cavity ergotropy and energy-basis coherence remain zero up to numerical precision, while the state converges to the Gibbs fixed point selected by the apparent detailed balance. We further estimate lower bounds on the cost of preparing the atomic populations and coherence, showing that the relevant advantage of phaseonium is a resource-preparation tradeoff rather than a cost-free enhancement over a thermal bath at the same temperature. Finally, we assess the finite-time performance of a two-cavity cascade with additive mechanical work accounting. Over the investigated coherence-phase range, the cascade produces approximately $47\%$--$52\%$ more power than the single-cavity engine while requiring only $65\%$--$68\%$ of the hot and cold phaseonium atoms needed by two independent engines, resulting in a $9\%$--$15\%$ enhancement of power per injected atom over a complete cycle.
\end{abstract}


\maketitle

Quantum thermodynamics extends classical thermodynamic principles to microscopic systems, where genuinely quantum resources such as coherence and correlations may affect heat, work, and irreversibility~\cite{streltsov_quantum-coherence_2017, binder_thermodynamics-in-the-quantum_2018, aimet_engineering_2023}. Coherence is especially subtle when it is placed in an environment rather than in the working medium. In this case the reservoir need not be thermal in the Gibbs sense, but it can still impose a well-defined excitation/dissipation ratio on the system.
This motivates the use of ``local'' or ``apparent'' temperatures, which characterize the stationary state and heat-flow direction of a system interacting with a non-thermal environment~\cite{alicki2_non-equilibrium_2015, latune_apparent_2019}.

Only coherences between degenerate levels can modify such an apparent temperature without introducing oscillating work-like terms in the system dynamics~\cite{latune_roles_2021, latune_apparent_2019, latune_quantum_2019, dag_multiatom_2016}.
Phaseonium~\cite{scully_from-lasers_1992} is one of the simplest systems where this mechanism appears.
It consists of a beam of three-level atoms in ${\rm V}$ or $\Lambda$ configuration, with coherence between the degenerate excited or ground states.
Its phase-dependent optical and thermodynamic properties have been studied from lasing without inversion to phase-controlled heat exchange~\cite{scully_from-lasers_1992,kozlov_ultrashort-pulses_2000, nguyen_coherently-tunable_2016, rathea_optical-nonlinearities_2021, scully_extracting-work_2002, amato_heating_2024}.

In the thermodynamic domain, phaseonium is studied because its internal coherence modifies the apparent temperature of a cavity.
Here we use this mechanism as a minimal and analytically transparent reservoir-engineering tool.
We ask how it performs in a concrete optomechanical Otto engine, where heat exchange, mechanical work, passive energy, and the cost of preparing the reservoir can be tracked explicitly.
Importantly, a phaseonium beam is not a freely supplied equilibrium bath.
It is a stream of coherently prepared atoms, and its thermodynamic role must be assessed against suitable reference preparations.

Optomechanical cavities~\cite{aspelmeyer_cavity_2014} provide a natural platform for this question.
Radiation pressure can move one mirror as a piston, while the cavity field is heated and cooled through engineered reservoirs.
Previous phaseonium-engine proposals often focused on idealized cycles or on comparisons with effective thermal bounds~\cite{quan_quantum-classical_2006, scully_extracting-work_2002}.
In contrast, we describe a finite-time Otto cycle whose isochores are generated by a collision model~\cite{ciccarello_quantum-collision_2022}, using the exact phaseonium-cavity dynamics derived in Ref.~\cite{amato_heating_2024}, and whose adiabatic strokes are implemented through the cavity length dependence of the field Hamiltonian.

Our contribution is threefold.
First, we connect microscopic phaseonium parameters to the detailed balance of the cavity dynamics and to mechanical work extracted by radiation pressure, and we show that the phaseonium-driven isochore charges the cavity passively: the cavity remains diagonal in the energy basis and its ergotropy is zero up to numerical precision.
Second, we compare the phaseonium engine with a thermal reference at the same apparent temperatures and with an incoherent reference at the same atomic populations, and we estimate lower bounds on the preparation cost of populations and coherence.
Third, we quantify the finite-time performance of a two-cavity cascade by accounting explicitly for the collision number, cycle duration, output power, and atomic consumption.
The resulting picture is a quantitative tradeoff between reservoir preparation, relaxation time, and power output.

\begin{figure*}[ht]
    \centering
    \begin{subfigure}{0.26\textwidth}
        \includegraphics[width=\textwidth]{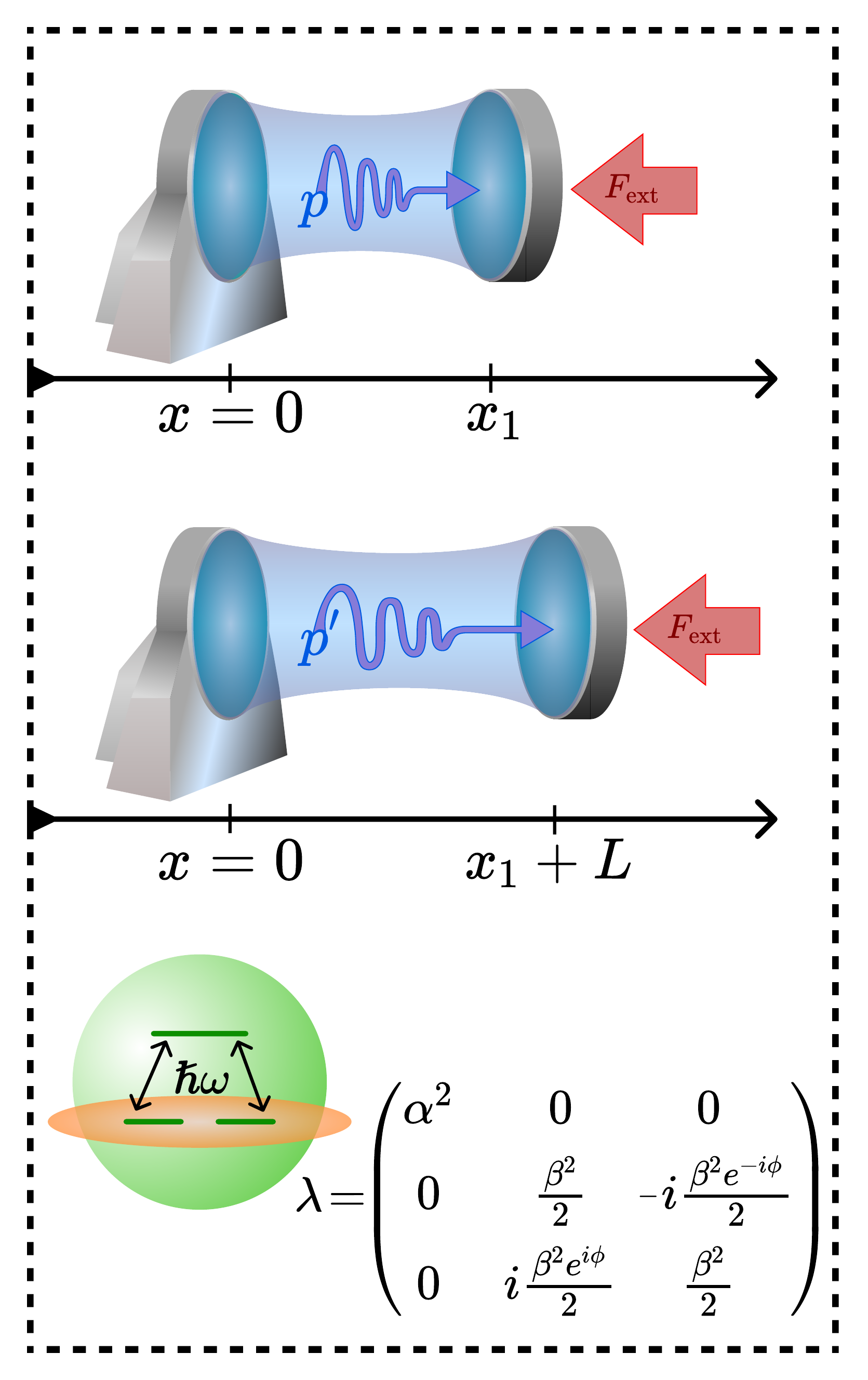}
        \caption{}
        \label{sfig:fluid-and-reservoir}
    \end{subfigure}
    \begin{subfigure}{0.44\textwidth}
        \includegraphics[width=\textwidth]{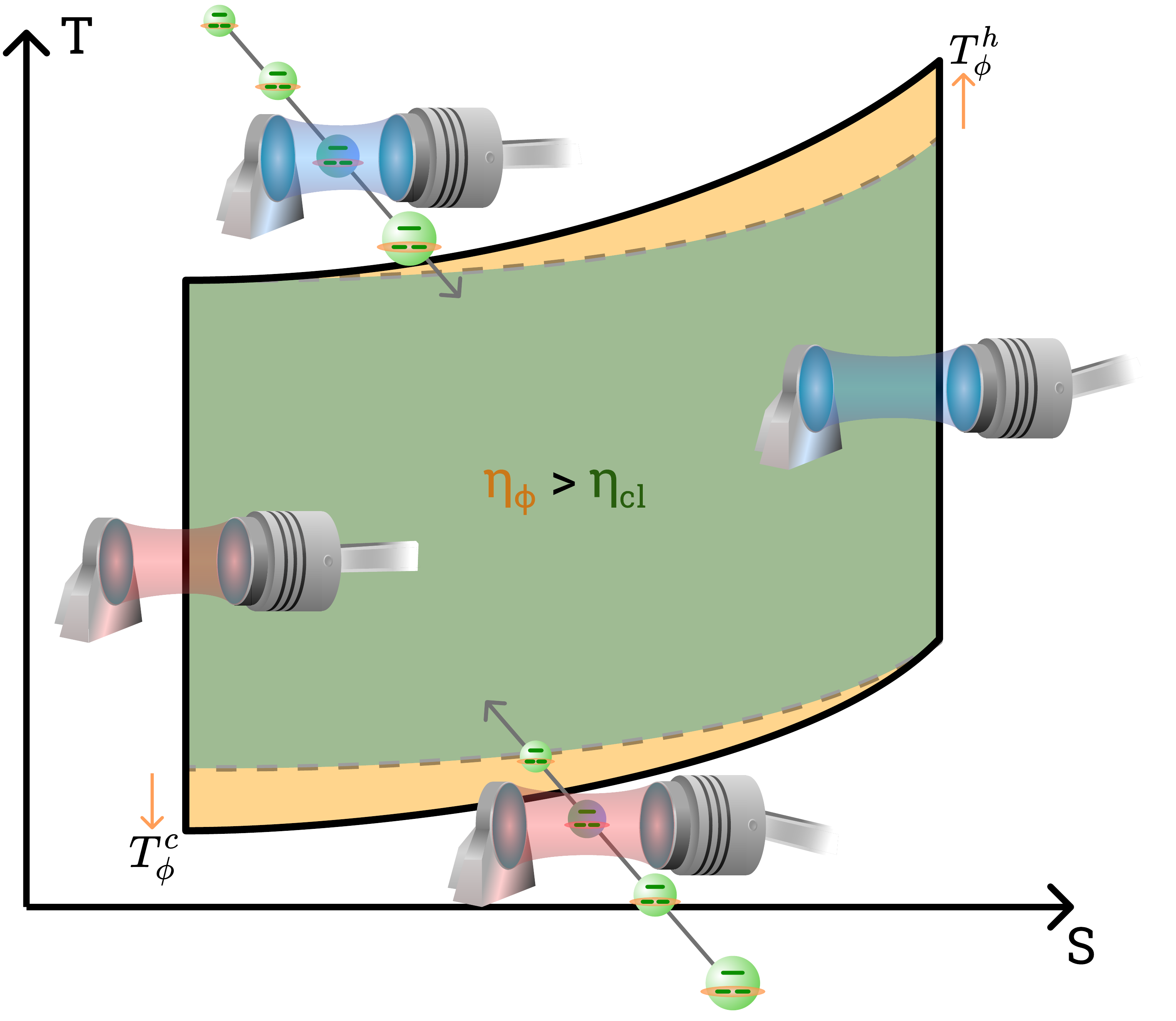}
        \caption{}
        \label{sfig:otto-cycle}
    \end{subfigure}
    \begin{subfigure}{0.26\textwidth}
        \includegraphics[width=\textwidth]{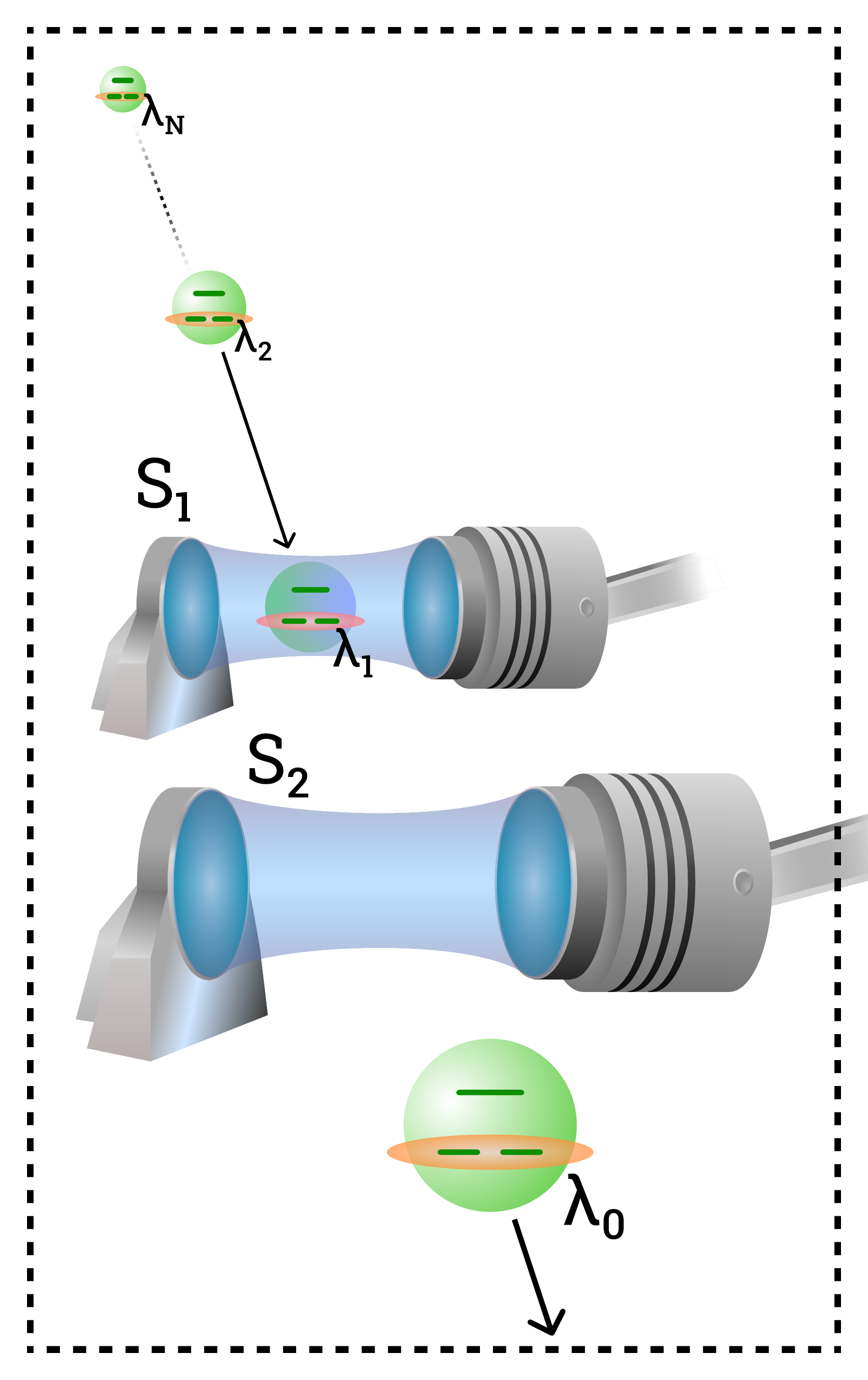}
        \caption{}
        \label{sfig:cascade-setup}
    \end{subfigure}
    \caption{Quantum heat engine architecture.
        (a) The working medium consists of a single-mode optical cavity with a movable end mirror, allowing the cavity length to vary from $x_1$ to $x_1 + L$ under the influence of radiation pressure. The cavity interacts with a stream of three-level atoms in $\Lambda$ configuration, characterized by partial coherence in the ground-state subspace and represented by the density matrix $\uplambda$. These atoms serve as a tunable quantum reservoir.
        (b) Thermodynamic Otto cycle plotted in the entropy--temperature (S--T) plane. The movable mirror is attached to a piston to transfer useful work to the outside environment. Isochoric strokes correspond to energy exchange with phaseonium atoms, while isentropic strokes are driven by mirror displacement. Adjusting the coherence phase $\phi$ of phaseonium atoms changes the detailed balance of the isochoric transformations and therefore the apparent temperatures that determine the Otto cycle.
        (c) Cascade setup: two identical cavities, each with a piston, are sequentially traversed by the same atomic beam, enabling heating or cooling via a shared phaseonium reservoir. The atomic flux is tuned so that each cavity contains at most one ancilla at a time, while different atoms may simultaneously occupy the two cavities. Thus, although every atom interacts sequentially with $S_1$ and then $S_2$, successive atoms can propagate through the cascade in a pipelined configuration.}
    \label{fig:setup}
\end{figure*}

Figure~\ref{fig:setup} shows the architecture considered throughout the paper.

\section{Model}
The system consists of two optical cavities in cascade configuration, interacting with a beam of three-level phaseonium atoms, as shown in Fig.~\ref{sfig:cascade-setup}.
The single-cavity engine is obtained by considering only the first cavity and is the reference setup for the thermodynamic analysis below.
For the cascade, each atom interacts first with $S_1$ and subsequently with $S_2$, while the following atom may already interact with $S_1$.
After the initial filling transient, one atom can therefore interact with each cavity simultaneously.
The two pistons are also driven synchronously during the adiabatic strokes.

The two cavity fields, $S_1$ and $S_2$, are modeled as standard single-mode harmonic oscillators~\cite{breuer_theory_2002}, with a controllable interaction with a stream of phaseonium atoms acting as engineered non-thermal reservoirs.
The coherent state that defines phaseonium atoms can be represented by the density operator $\uplambda$ presented in Fig.~\ref{sfig:fluid-and-reservoir}, with the normalization condition $\abs{\alpha}^2~=~1-\abs{\beta}^2$.
We choose a resonant coupling for the cavity and atomic frequencies and use the interaction picture to leave the free evolution of both the cavity and the bath out of the analysis.
We use the collision model to study the thermalization dynamics of the systems interacting with the phaseonium.
This is a thermalization process toward an apparent temperature given by the phaseonium atoms' coherence and population~\cite{amato_heating_2024}:
\begin{equation}\label{eq:steady-state-temperature}
    k_B T_\phi = \frac{\hbar\omega}{\ln{\left(\frac{\beta^2( 1 -\sin\phi)}{2\alpha^2}\right)}} \,,
\end{equation}

For comparison, the master equation for a cavity interacting with a thermal (diagonal) three-level atom in $\Lambda$ configuration---without quantum coherence---leads to a stable temperature for the cavity
\begin{equation}\label{def:classic-temperature}
    k_B T_{cl} = \frac{\hbar\omega}{\ln\left({\frac{\beta^2}{2\alpha^2}}\right)} \,.
\end{equation}

\begin{figure}[h!]
    \centering
    \includegraphics[width=\linewidth]{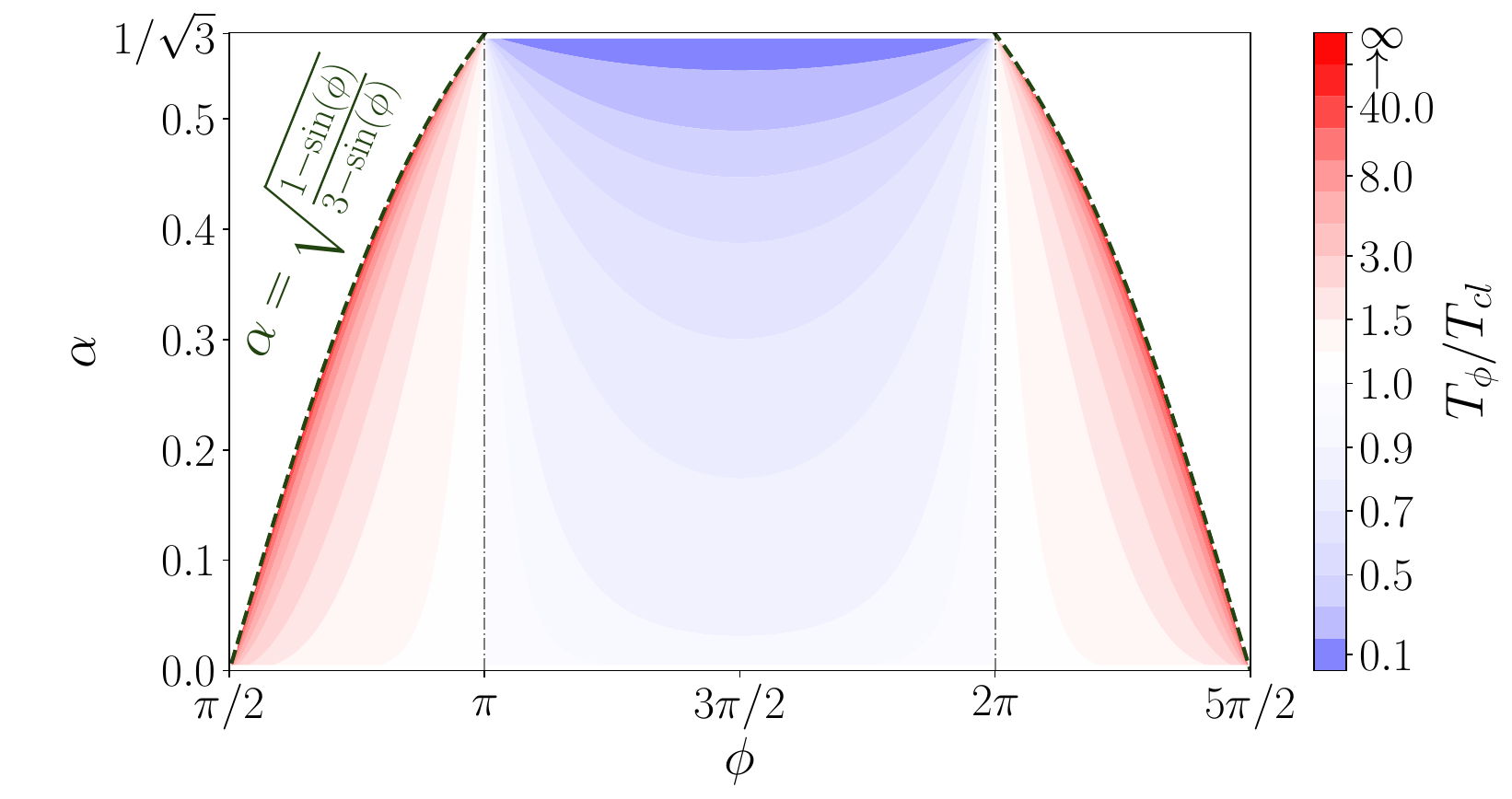}
    \caption{Ratio between the apparent temperatures induced in the cavity by phaseonium atoms, $T_\phi$, and by diagonal $\Lambda$ atoms, $T_\text{cl}$. The classical temperature $T$ is defined only for $\alpha<\sqrt{1/3}$, while phaseonium temperature $T_\phi$ is defined for $\alpha<\sqrt{(1-\sin\phi)/(3-\sin\phi)}$, limited by the dashed black line represented in the plot. For $\phi=k\pi$ and $k\in\mathbb{Z}_0$, along the dash-dotted grey lines, the two temperatures are the same.}\label{fig:quantum-to-classic-T}
\end{figure}

Figure~\ref{fig:quantum-to-classic-T} shows the quantum to classical ratio of temperatures $T_\phi/T_{cl}$.
From this figure, it is clear that for $2k\pi<\phi<(2k+1)\pi$ and $k\in\mathbb{Z}_0$, given a certain $\alpha$ within the boundaries of Eq.~\eqref{eq:steady-state-temperature} and Eq.~\eqref{def:classic-temperature}, coherence increases the stable temperature of the pumped cavity with respect to the diagonal atomic reference.
In contrast, for $(2k-1)\pi<\phi<2k\pi$, the final temperature of the pumped cavity is lower than in the diagonal reference, meaning that coherence changes the balance between emission and absorption in the opposite direction.
The two temperatures are the same for $\phi=k\pi$ , when coherence is purely imaginary.
\medskip

\subsection{Thermodynamic interpretation of the phaseonium isochore}\label{sec:thermodynamic-interpretation}
The temperature in Eq.~\eqref{eq:steady-state-temperature} should be understood operationally. The phaseonium atoms are not themselves in a Gibbs state at $T_\phi$. Rather, the collision map representing the cavity-phaseonium interactions fixes the ratio between the cavity excitation and de-excitation rates,
\begin{equation}\label{eq:detailed-balance-phaseonium}
    \frac{\Gamma_\uparrow(\ad)}{\Gamma_\downarrow( \a )}
    =\frac{2\alpha^2}{\beta^2(1-\sin\phi)}
    =\exp\left[-\frac{\hbar\omega}{k_B T_\phi}\right] .
\end{equation}
The apparent temperature is therefore a detailed-balance parameter of the reduced cavity dynamics. In this sense, phaseonium coherence acts directly on the balance of gain and loss processes of the working fluid.

This is a property of the fixed-point steady state of the isochore map, verified analytically using the collision-model framework.
What follows is a complete thermodynamic characterization of the whole out-of-equilibrium evolution of the cavity state under the interaction with phaseonium atoms during the isochoric stroke, starting from the kind of energy delivered to the cavity.
For a state $\rho_S$ of the cavity with Hamiltonian $H_S$, the ergotropy is the maximum work extractable by a cyclic unitary acting on the cavity alone~\cite{allahverdyan_maximal_2004}. It can be written as
\begin{equation}\label{eq:ergotropy}
    \mathcal{E}_S(\rho_S)=\Tr[H_S\rho_S]-\Tr[H_S\rho_S^{\rm pas}],
\end{equation}
where $\rho_S^{\rm pas}$ is the passive state obtained by rearranging the eigenvalues of $\rho_S$ in decreasing order over increasing energy levels~\cite{pusz_passive_1978,lenard_thermodynamical_1978}. If $\mathcal{E}_S=0$, the internal energy of the cavity cannot be extracted as work without changing the Hamiltonian or coupling to another system.

\begin{figure}[h]
    \centering
    \includegraphics[width=0.8\linewidth]{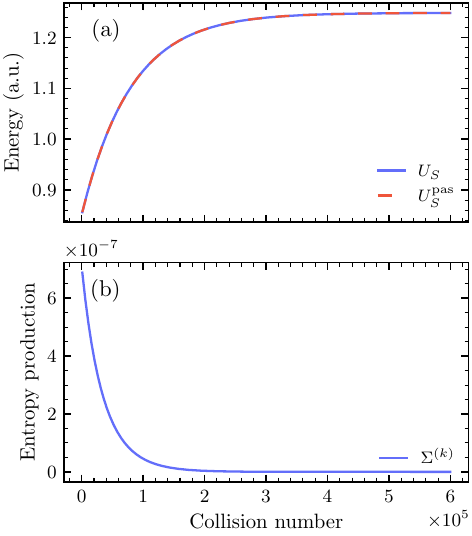}
    \caption{Cavity energy and entropy production of collisions during the isochoric strokes. (a) The internal energy $U_S$ exchanged with the coherent phaseonium atoms is all stored as passive energy $U_S^{\rm pas}$, thus the two lines coincide. The coincidence of the two curves shows that the cavity ergotropy remains zero during the isochoric stroke. This establishes passivity of the evolving state. (b) Single-collision entropy production $\Sigma^{(k)}$ of the reduced cavity dynamics. Each point refers to the $k$th collision. Its positivity verifies the discrete Clausius inequality associated with the effective temperature $T_\phi$, while its decay to zero shows that irreversibility vanishes as the cavity approaches the Gibbs fixed point.}
    \label{fig:passive-energy}
\end{figure}

In our engine, the cavity starts each isochore from a thermal and therefore passive state.
The phaseonium collision map preserves diagonality in the cavity energy basis and drives the
oscillator toward the Gibbs fixed point associated with Eq.~\eqref{eq:detailed-balance-phaseonium}.
Numerically, we find that the
ergotropy $\mathcal{E}_S$ remains zero up to numerical precision throughout the isochoric
stroke.
Hence, the energy exchanged with the atomic beam is stored as passive energy rather
than as locally extractable work, as shown in Fig.~\ref{fig:passive-energy}a.

Let $\Phi_\phi$ denote the completely positive trace-preserving map associated with a single cavity--atom collision, and, using Eq.~\ref{eq:steady-state-temperature}, define its invariant Gibbs state as
\begin{equation}\label{eq:steady_state}
    \tau_\phi
    =
    \frac{e^{-H_S/(k_B T_\phi)}}{Z_\phi},
\end{equation}
satisfying $\Phi_\phi(\tau_\phi)=\tau_\phi$. For $\rho_{n+1}=\Phi_\phi(\rho_n)$,
the contractivity of the quantum relative entropy $S(\varrho\,\Vert\,\varrho')$ implies
\begin{equation}
    \Sigma^{(k)}
    \equiv
    S \!\left(\rho_k\middle\|\tau_\phi\right)
    -
    S\!\left(\rho_{k+1}\middle\|\tau_\phi\right)
    \geq 0.
\end{equation}
This decrease of the relative entropy with respect to the stationary state defines the coarse-grained entropy production $\Sigma^{(k)}$ associated with the $k$th collision~\cite{landi_irreversible_2021,ciccarello_quantum-collision_2022}.
Since the cavity Hamiltonian remains constant during the isochoric stroke, the energy variation
\begin{equation}
    Q_k
    =
    \Tr\!\left[
        H_S\left(\rho_{k+1}-\rho_k\right)
        \right]
\end{equation}
is heat-like. Moreover, using the Gibbs form of $\tau_\phi$, the entropy production can be rewritten as
\begin{equation}
    \Sigma^{(k)}
    =
    \Delta S_S^{(k)}
    -
    \frac{Q_k}{(k_B T_\phi)}
    \geq 0,
\end{equation}
where
\begin{equation}
    \Delta S_S^{(k)}
    =
    S(\rho_{k+1})-S(\rho_k).
\end{equation}
Therefore, the phaseonium-induced temperature $T_\phi$ is not merely a parameter characterizing the stationary cavity occupation.
It is the temperature entering the entropy-flow contribution for which the reduced cavity dynamics satisfies a discrete Clausius inequality.
In this operational sense, $T_\phi$ is a thermodynamically consistent effective temperature.
We stress, however, that $\Sigma^{(k)}$ describes only the coarse-grained entropy production of the reduced cavity dynamics and does not include the work required to prepare the nonthermal phaseonium atoms, the residual thermodynamic resources carried by the outgoing atoms, or the correlations generated during each collision.

Passivity alone, however, does not imply that a state is thermal~\cite{perarnau_most_2015}, and neither that the transformation is reversible.
To characterize separately the absence of energy-basis coherence, we compute the relative entropy of coherence
\begin{equation}
    C_{\mathrm{rel.ent.}}(\rho_S)
    =
    S\!\left(\mathcal{D}[\rho_S]\right)-S(\rho_S),
\end{equation}
where $\mathcal{D}[\rho_S]$ denotes complete dephasing in the cavity energy basis.
This quantity remains zero up to numerical precision, confirming that no energy-basis
coherence is generated during the isochore.
The cavity state is therefore diagonal and passive
throughout the evolution and converges to the Gibbs fixed point selected by the phaseonium detailed balance.

Although the cavity remains passive during the isochoric stroke, each phaseonium atom can create correlations and lose part of its initial coherence during its collision with the cavity.
Let $\uplambda_{\mathrm{in}}^k$ denote the common incoming atomic state and let $\uplambda_{\mathrm{out}}^k$ denote the state of the atom leaving the cavity after the $k$th collision.
First, we quantify the entropy production associated with the $k$th collision through the quantum relative entropy between the joint states after and before the interaction.
The incoming cavity and atom are initially uncorrelated,
\begin{equation}
    \rho_S^{(k-1)}\otimes \uplambda_\text{in}^k,
\end{equation}
and evolve unitarily to
\begin{equation}
    \sigma^{(k)}
    =
    U_{SA}\left(
    \rho_S^{(k-1)}\otimes \uplambda_\text{in}^k
    \right) U_{SA}^{\dagger}.
\end{equation}

\begin{figure*}[t!]
    \centering
    \includegraphics[width=0.4\linewidth]{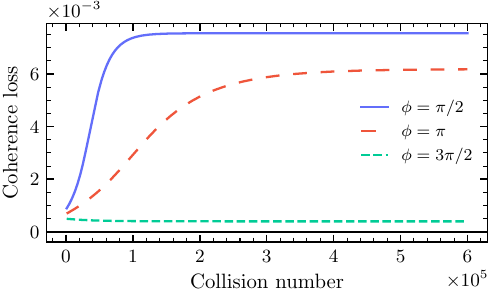}\hspace{0.1\linewidth}
    \includegraphics[width=0.4\linewidth]{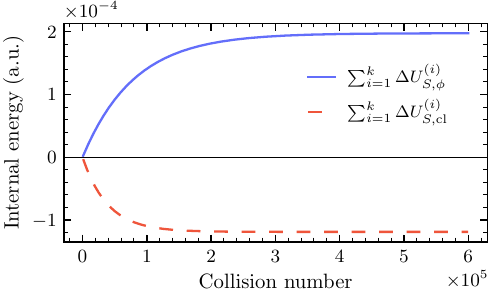}
    \caption{Thermodynamic diagnostics of the phaseonium-driven isochore. Left: single-collision coherence loss $\Delta C_A^{(k)} =C_{\mathrm{rel.ent.}}(\uplambda_{\mathrm{in}}^k) -C_{\mathrm{rel.ent.}}(\uplambda_{\mathrm{out}}^k)$ for the atom involved in the $k$th collision, shown for different phaseonium phases. The plotted quantity measures how much relative entropy of coherence an individual atom loses while interacting with the cavity at the corresponding stage of the thermalization process. The loss approaches a phase-dependent stationary value as the cavity approaches its fixed point. Right: cumulative internal-energy change of the cavity for phaseonium atoms and for an incoherent same-population reference. The phaseonium beam can increase the passive energy of the cavity while the corresponding incoherent preparation may cool it, showing that the coherence changes the detailed balance generated by the same populations. In the same simulations, the cavity ergotropy and relative entropy of coherence remain zero up to numerical precision. The field therefore remains diagonal and passive while approaching the Gibbs fixed point.}\label{fig:phaseonium-thermo}
\end{figure*}

Second, we define the single-collision coherence loss as
\begin{equation}
    \Delta C_A^{(k)}
    \equiv
    C_{\mathrm{rel.ent.}}\!\left(\uplambda_\text{in}^k\right)
    -
    C_{\mathrm{rel.ent.}}\!\left(\uplambda_\text{out}^k\right).
    \label{eq:atomic-coherence-loss}
\end{equation}
A positive value therefore means that the outgoing atom contains less relative entropy of coherence than it had before the collision. This quantity is not cumulative: each point in the left panel of Fig.~\ref{fig:phaseonium-thermo} is the coherence lost by the individual atom involved in that collision.
We repeat the isochoric thermalization for different values of the phase $\phi$ and evaluate
$\Delta C_A^{(k)}$ at the sampled collision numbers. As the cavity approaches its stationary
state, the single-collision loss approaches a phase-dependent stationary value.
The right panel compares the
cumulative cavity-energy variation generated by the phaseonium beam
with that generated by its fully dephased counterpart at the same
atomic populations. In the displayed regime, the coherent beam heats
the cavity whereas the dephased beam cools it. The field is therefore not charged as a quantum battery. Its energy gain is passive energy determined by the apparent temperature given by the detailed balance.

\subsection{Engine cycle and work accounting}
The simplest cycle to implement with this working medium is an Otto cycle, represented in Fig.~\ref{sfig:otto-cycle}, where energy exchange with the atomic beams occurs during the two isochoric strokes and mechanical work is associated with the two adiabatic strokes. During an isochore, the cavity length is fixed and the Hamiltonian is constant. We therefore identify the heat-like energy exchanged with the atoms as
\begin{equation}\label{eq:heat-isochore}
    Q = \Delta U_S = \Tr[H_S(\rho_S^{\rm f}-\rho_S^{\rm i})] .
\end{equation}
During an adiabatic stroke the state is transported while the cavity frequency changes through the length dependence $\omega(t)\propto 1/L(t)$.

For the adiabatic strokes, we use the radiation-pressure operator of a cavity mode~\cite{aspelmeyer_cavity_2014,law_interaction_1995,tejero_atom-doped_2024}, to the order $\mathcal{O}({(\hat{a}^\dagger\hat{a})}^2)$,
\begin{equation}\label{def:radiation-pressure}
    \pi(t) \equiv \frac{\omega(t)}{2V(t)}\left( \ad\a + \a\ad - \a\a e^{-i2\omega t} - \ad\ad e^{i2\omega t} \right) .
\end{equation}
Neglecting the rotating terms, the pressure is $p(t)=\Tr[\pi(t)\rho(t)]=\omega(t)(\langle\hat n\rangle+1/2)/V(t)$. For two independent pistons, the global pressure operator is additive,
\begin{equation}\label{eq:additive-pressure}
    \Pi(t)=\pi_{S_1}(t)\otimes\idd_{S_2}+\idd_{S_1}\otimes\pi_{S_2}(t)\,.
\end{equation}

For a single cavity pushing a piston of surface $S$, the mechanical work reads
\begin{equation}\label{def:mechanical-work}
    W^m = \int_{L_1}^{L_2}p(L)S\,dL .
\end{equation}
Using $dL/L=-d\omega/\omega$ and neglecting the zero-point contribution, this becomes
\begin{equation}\label{eq:mechanical-work-frequency}
    W^m \simeq -\int_{\omega_1}^{\omega_2}\langle \hat n\rangle d\omega .
\end{equation}
This classical expression of work can be compared with the pure quantum equation given by Alicki~\cite{alicki_quantum_1979}, which stems directly from the first principle of thermodynamics:
\begin{equation}\label{def:alicki-definitions}
    \frac{d}{dt}\ev{H(t)}=
    \underbrace{\Tr\left[\rho(t)\frac{dH(t)}{dt}\right]}_{\dot W^{\rm Al}}
    +
    \underbrace{\Tr\left[\frac{d\rho}{dt}H(t)\right]}_{\dot Q^{\rm Al}} .
\end{equation}
Our sign convention is that extracted work is negative in the energy balance of the cavity. We therefore quote positive work output as $W_{\rm out}=-W_{\rm net}$ and define the bare cycle efficiency as
\begin{equation}\label{eq:bare-efficiency}
    \eta_{\rm cyc}=\frac{W_{\rm out}}{Q_{\rm in}}=-\frac{W_{\rm net}}{Q_{\rm in}} .
\end{equation}
We note that this bare efficiency characterizes the working-fluid cycle only, and does not take into account the cost of preparing the reservoirs.

\subsection{Reservoir preparation cost}
To compare phaseonium with thermal or incoherent references, we estimate the minimum work required to prepare the incoming atoms from a reference thermal state at temperature $T_0$. For an atomic state $\uplambda$ and Hamiltonian $H_\uplambda$, the non-equilibrium free energy is
\begin{equation}\label{eq:noneq-free-energy}
    F_{T_0}(\uplambda)=\Tr[H_\uplambda\uplambda]-k_B T_0 S(\uplambda) .
\end{equation}
The preparation work is lower bounded by the corresponding non-equilibrium free-energy difference~\cite{brandao_resource_2013, skrzypczyk_work_2014},
\begin{equation}\label{eq:prep-lower-bound}
    W_{\rm prep}^{\rm lb}(T_0)=F_{T_0}(\uplambda)-F_{T_0}(\tau_0),
\end{equation}
where $\tau_0=\exp[-H_\uplambda/(k_BT_0)]/Z_0$ is the reference Gibbs state. We separate this into a population contribution and a coherence contribution by introducing the dephased state $\mathcal{D}[\uplambda]$ in the atomic energy basis,
\begin{align}
    W_{\rm pop}^{\rm lb}(T_0) & =F_{T_0}(\mathcal{D}[\uplambda])-F_{T_0}(\tau_0),\label{eq:population-cost}   \\
    W_{\rm coh}^{\rm lb}(T_0) & =F_{T_0}(\uplambda)-F_{T_0}(\mathcal{D}[\uplambda]).\label{eq:coherence-cost}
\end{align}
Since the phaseonium coherence lies inside a degenerate subspace, it does not change the mean atomic energy. Its lower-bound cost is therefore entropic,
\begin{equation}\label{eq:coh-cost-crel}
    W_{\rm coh}^{\rm lb}(T_0)=k_B T_0 C_{\rm rel.ent.}(\uplambda),
\end{equation}
where $C_{\rm rel.ent.}(\uplambda)=S(\mathcal{D}[\uplambda])-S(\uplambda)$ is the relative entropy of coherence~\cite{baumgratz_quantifying_2014,lostaglio_description_2015}. This decomposition is a lower bound, but it is sufficient to show that coherence-assisted operation must be evaluated as a resource-preparation tradeoff.

The per-atom preparation cost is not, by itself, a complete measure of the resources consumed by the reservoir. Even when thermal and phaseonium atoms drive the cavity toward the same apparent temperature, they can generate different relaxation rates and therefore require different numbers of collisions to complete an isochoric stroke. The apparent temperature fixes only the detailed-balance ratio in Eq.~\eqref{eq:detailed-balance-phaseonium}, whereas the thermalization speed is controlled by the absolute relaxation gap
\begin{equation}\label{eq:relaxation-gap}
    \kappa_r
    =
    \Gamma_{\downarrow,r}
    -
    \Gamma_{\uparrow,r},
    \qquad r=h,c.
\end{equation}

\begin{figure}[h]
    \centering
    \includegraphics[width=0.7\linewidth]{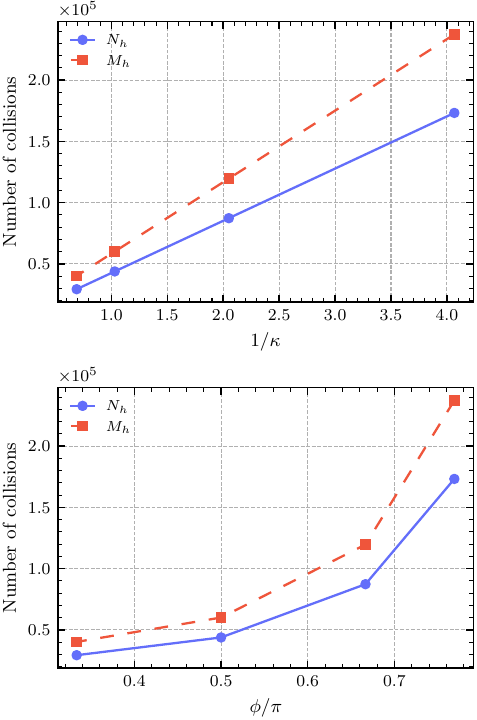}
    \caption{Minimum hot-isochore collision numbers for the single-cavity engine, $N_h$, and the cascade, $M_h$. Both scale approximately linearly with the inverse relaxation gap $1/\kappa_h$, while $N_h<M_h<2N_h$.}
    \label{fig:cascade-n-collisions}
\end{figure}

For a fixed initial state and convergence tolerance, the minimum collision number consequently scales approximately as
\begin{equation}\label{eq:collision-gap-scaling}
    N_r^{\rm th}
    \propto
    \frac{1}{\kappa_r}.
\end{equation}
This behavior is verified explicitly for both the single-cavity and cascade protocols in Fig.~\ref{fig:cascade-n-collisions}, using the hot isochore as reference. Thus, coherence affects not only the stationary apparent temperature but also the atomic flux and time required to realize the corresponding thermodynamic stroke.

Therefore, we include both the lower-bound preparation cost of each incoming atom and the
number of atoms consumed during the hot and cold isochoric strokes. For each preparation
route $x\in\{\phi,\mathrm{th}\}$, we define the collision-weighted total preparation-cost lower
bound over one complete cycle as
\begin{equation}
    C_{x}^{\mathrm{lb}}(T_0)
    =
    N_{h,x}\,w_{h,x}^{\mathrm{lb}}(T_0)
    +
    N_{c,x}\,w_{c,x}^{\mathrm{lb}}(T_0),
    \label{eq:total-reservoir-cost}
\end{equation}
where $N_{h,x}$ and $N_{c,x}$ are the minimum numbers of atoms required to complete the
hot and cold isochores, respectively, and $w_{r,x}^{\mathrm{lb}}$ is the corresponding per-atom
preparation-cost lower bound for reservoir $r=h,c$. For the phaseonium route,
$w_{r,\phi}^{\mathrm{lb}}$ includes both the population and coherence contributions, whereas
the thermal route contains only the population-preparation contribution required to reproduce
the same apparent temperature.

\begin{figure}
    \includegraphics[width=0.45\textwidth]{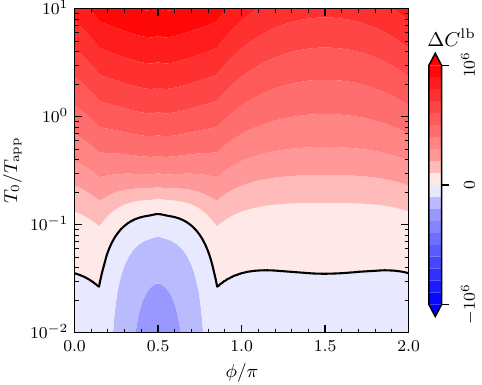}
    \caption{Difference between the collision-weighted lower bounds on the total preparation costs of the phaseonium and thermal reservoir routes, $\Delta C^{\mathrm{lb}}=C_{\phi}^{\mathrm{lb}}-C_{\mathrm{th}}^{\mathrm{lb}}$. Both $C_{\phi}^{\mathrm{lb}}$ and $C_{\mathrm{th}}^{\mathrm{lb}}$ include the hot- and cold-reservoir contributions and the corresponding minimum collision numbers required to complete the isochores. The thermal reservoirs are chosen to reproduce the same apparent temperatures as the phaseonium reservoirs. The black contour marks $\Delta C^{\mathrm{lb}}=0$. In the region where $\Delta C^{\mathrm{lb}}<0$, the phaseonium reservoir has the lower preparation-cost bound; where $\Delta C^{\mathrm{lb}}>0$, the thermal reservoir preparation has the lower bound.}
    \label{fig:reservoir-cost-difference}
\end{figure}

The quantity plotted in Fig.~\ref{fig:reservoir-cost-difference} is
\begin{equation}
    \Delta C^{\mathrm{lb}}(T_0,\phi)
    \equiv
    C_{\phi}^{\mathrm{lb}}(T_0,\phi)
    -
    C_{\mathrm{th}}^{\mathrm{lb}}(T_0,\phi).
    \label{eq:cost-difference}
\end{equation}
Thus, $\Delta C^{\mathrm{lb}}<0$ means that the phaseonium preparation route has the lower
collision-weighted preparation-cost bound, while $\Delta C^{\mathrm{lb}}>0$ means that the
thermal preparation route has the lower bound. These quantities are thermodynamic lower
bounds and do not include implementation-dependent experimental costs.

\section{Results}

\subsection{Phaseonium Engine}
We first compare three engines that follow the same optomechanical protocol. The first is a thermal reference, where the hot and cold reservoirs are ordinary thermal reservoirs chosen to reproduce the same apparent temperatures $T_\phi^h$ and $T_\phi^c$ induced by the phaseonium beams. The second is an incoherent atomic reference with the same populations as the phaseonium atoms, but with the ground-state coherence removed. The third is the phaseonium engine.

This comparison separates two distinct effects of the atomic preparation. At fixed apparent
temperatures, the phaseonium and thermal-reference reservoirs select the same stationary Gibbs
states. Consequently, when both isochoric strokes are continued until convergence, they generate
the same fully thermalized bare Otto cycle. This stationary-state equivalence does not imply
identical finite-time dynamics: the two preparations may have different absolute excitation and
de-excitation rates, relaxation gaps, collision numbers, cycle durations, and powers. At fixed
atomic populations, instead, removing the coherence changes the detailed-balance ratio and
therefore produces a different Otto cycle. Phaseonium is useful in this second sense: it changes the apparent hot and cold temperatures generated by a given population preparation.

\begin{figure}[h!]
    \centering
    \includegraphics[width=0.7\linewidth]{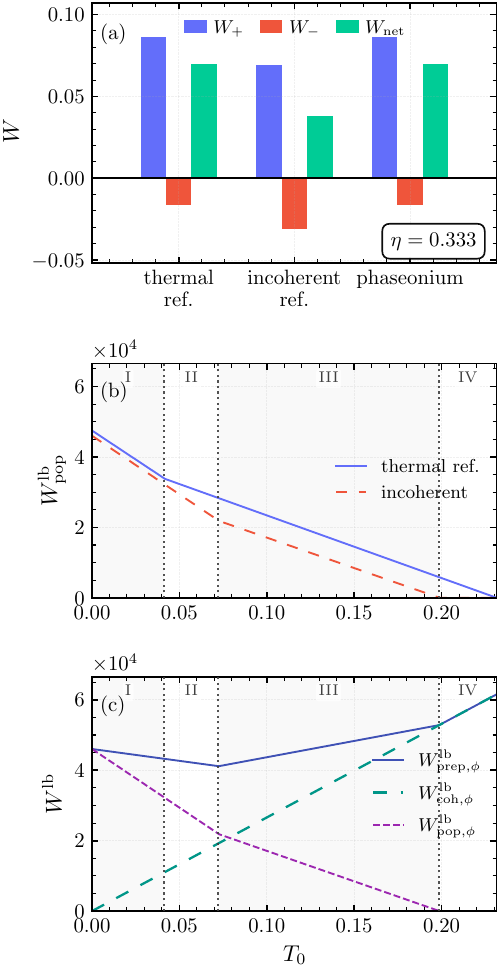}
    \caption{Bare cycle performance and preparation-cost lower bounds. (a) Work contributions for three reservoir preparations: a thermal reference at the same apparent temperatures $\uplambda_\text{th}$, an incoherent same-population atomic beam $\mathcal{D}[\uplambda]$, and the phaseonium beam $\uplambda$. Once the isochores have reached their stationary states, the thermal same-$T_\phi$ and phaseonium preparations produce the same fully thermalized cycle, with the same work output and absorbed heat. The incoherent same-population reference gives a smaller net work output because it works between different temperatures, closer together. At the same time, it absorbs less heat. Therefore, the efficiency for these three reservoirs is the same. (b) Lower bound on the population-preparation work as a function of the reference temperature $T_0$, comparing the thermal reference and the phaseonium population preparation (same as the incoherent atoms). (c) Decomposition of the phaseonium preparation lower bound into population and coherence contributions. Creating coherence inside the degenerate subspace has no energetic contribution, but it has an entropic free-energy cost proportional to $T_0 C_{\rm rel.ent.}$. The figure shows a resource tradeoff between populations and coherence preparation for the same engine efficiency.}\label{fig:engine-costs}
\end{figure}

The work data are shown in Fig.~\ref{fig:engine-costs}(a). The thermal same-$T_\phi$ reference and the phaseonium engine give the same bare cycle efficiency, $\eta_{\mathrm{cyc}}$, and the same net work once the hot and cold isochores have reached their stationary states.
For comparison, the bare cycle efficiency computed from Eq.~\eqref{eq:bare-efficiency} is below the Curzon--Ahlborn efficiency $\eta_\text{CA}$ benchmark~\cite{curzon_efficiency_1975,cangemi_quantum_2024},
\begin{equation}
    \eta_\text{cyc}\approx 0.333 < 1 - \sqrt{\frac{T_\phi^c}{T_\phi^h}}\approx 0.388
\end{equation}

The incoherent same-population reference behaves differently. Since the coherence is removed while the populations are kept fixed, the gain and loss rates no longer satisfy the same detailed balance.
The apparent temperatures are shifted and the cycle area is reduced, with a smaller net work output.

For a more precise calculation of efficiency, we must account for the cost of preparing the reservoir.
The preparation-cost lower bounds for the engines considered in this section are shown in Fig.~\ref{fig:engine-costs}, panels~b and~c. The cost of preparing populations decreases with the temperature of the reference state, while the coherence contribution grows linearly with $T_0$ because the coherent levels are degenerate and the cost is purely entropic.

This comparison emphasize the operational role of coherence, that is a control parameter for the detailed balance of the cavity after thermalization.
For the same atomic populations, coherence can enhance or hinder the energy exchange from the reservoir to the system.

The comparison therefore identifies a trade-off between population engineering and coherence preparation. In the region where $\Delta C^{\mathrm{lb}}<0$, the phaseonium route has a lower preparation cost bound than the thermal reference.
Whether this translates into an experimental advantage depends on implementation-specific preparation costs and efficiencies.

\subsection{Finite-time cascade performance}\label{sec:cascade-performance}

The same collision-model framework naturally allows one phaseonium beam to drive more than one working medium.
Therefore, we compare two architectures.
Setup A consists of two identical single-cavity engines operated independently in parallel, each supplied by separate but identically prepared hot and cold phaseonium beams.
Setup B is the two-cavity cascade, shown in Fig.~\ref{sfig:cascade-setup}, in which each atom interacts first with $S_1$ and then with $S_2$, mediating a unidirectional flow of information from the upstream to the downstream cavity.
Mechanical work remains additive because the two cavities act on independent pistons,
\begin{equation}\label{eq:cascade-additive-work}
    W_{2}^{\rm out}
    =
    W_{S_1}^{\rm out}
    +
    W_{S_2}^{\rm out}\,.
\end{equation}

Although $W_{2}^{\rm out}$ will be greater than the work output of one cavity alone, the downstream cavity in setup B requires more collisions to reach its target thermal state.
We therefore determine the duration of each isochoric stroke dynamically.
The evolution is stopped when the last cavity reaches the corresponding stationary temperature within the same prescribed numerical tolerance.

We denote by $N_r$ the minimum number of atoms required by the single-cavity engine during isochore $r=h,c$, and by $M_r$ the corresponding number for the two-cavity cascade.
Also, $N=N_h+N_c$ and $M=M_h+M_c$ denote the total number of collisions for a full cycle.
One cascade collision step corresponds to the injection of one atom that traverses both cavities.
Because successive atoms propagate through the two cavities in a pipeline, the cycle durations are
\begin{align}
    \tau_A
     & =
    (N_h+N_c)\Delta t_{\rm col}
    +
    \tau_{\rm ad},
    \label{eq:single-cycle-time}
    \\
    \tau_B
     & =
    (M_h+M_c)\Delta t_{\rm col}
    +
    \tau_{\rm ad},
    \label{eq:cascade-cycle-time}
\end{align}
where $\tau_{\rm ad}$ is the total duration of the synchronized adiabatic strokes, same for both architectures.
The corresponding output powers are
\begin{equation}\label{eq:cascade-powers}
    P_1 = \frac{W_1^{\rm out}}{\tau_1},
    \qquad
    P_B = \frac{W_2^{\rm out}}{\tau_2},
\end{equation}
for one single-cavity engine and the cascade, respectively.
Clearly $P_A=2P_1$.

Figure~\ref{fig:cascade-n-collisions} shows the collision numbers required for the hot-stroke thermalization.
Both $N_h$ and $M_h$ increase approximately linearly with $1/\kappa_h$, consistently with Eq.~\eqref{eq:collision-gap-scaling}.
The cascade always requires more atoms than one cavity but substantially fewer than two independently thermalized cavities,
\begin{equation}\label{eq:atom-number-ordering}
    N_h<M_h<2N_h.
\end{equation}
The downstream cavity therefore increases the cycle duration, but the same atomic beam is reused before being discarded.

\begin{figure*}[t!]
    \centering
    \begin{subfigure}{0.4\textwidth}
        \includegraphics[width=\linewidth]{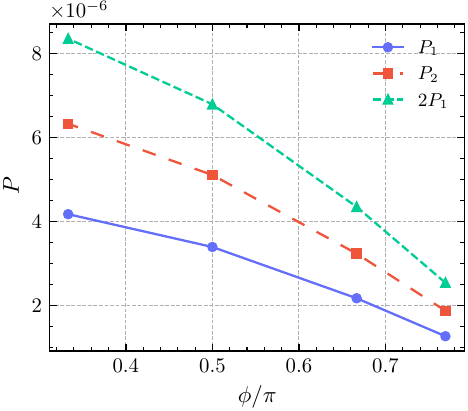}
        \caption{}
        \label{sfig:cascade-power}
    \end{subfigure}
    \hfill
    \begin{subfigure}{0.4\textwidth}
        \includegraphics[width=\linewidth]{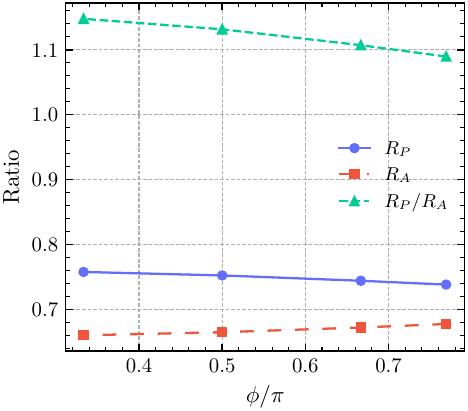}
        \caption{}
        \label{sfig:cascade-ratios}
    \end{subfigure}
    \caption{
        Finite-time performance of the two-cavity cascade.
        (a) Output powers of the single-cavity engine, $P_1$, and the cascade, $P_B$, as functions of the hot coherence phase. The curve $2P_1\equiv P_A$ is the power of two independent copies of the single-cavity engine. The cascade consistently satisfies $P_1<P_B<2P_1$.
        (b) Retained-power ratio $R_P=P_B/P_A$, atomic-resource ratio $R_a=M/(2N)$, and their quotient. Since $R_P/R_a>1$, the cascade delivers more power per injected phaseonium atom than two single-cavity independent engines.
    }
    \label{fig:cascade-performance}
\end{figure*}

The numerical results of the coherence-phase scan are reported in
Table~\ref{tab:cascade-scan}. At every point, the cascade produces
more power than the single-cavity engine, $P_B>P_1$, although it does
not reach the power $P_A$ of two independent engines, as shown in Fig.~\ref{sfig:cascade-power}.
To compare this power gain with the atomic resources consumed, as in Fig.~\ref{sfig:cascade-ratios}, we introduce
\begin{equation}\label{eq:cascade-resource-ratios}
    R_P
    \equiv
    \frac{P_B}{P_A},
    \qquad
    R_a
    \equiv
    \frac{M}{2N}.
\end{equation}
Here, $R_P$ is the fraction of the power of two independent single-cavity engines retained by the cascade, while $R_a$ is the corresponding fraction of the total phaseonium atoms consumed during the hot and cold isochores.
The ratio of powers normalized by number of required atoms is
\begin{equation}\label{eq:power-per-atom-ratio}
    \frac{P_B/M}{P_A/(2N)} = \frac{R_P}{R_a}\,.
\end{equation}
Throughout the scan, $R_P/R_a>1$, showing that the retained fraction of power $R_P$ is systematically larger than the corresponding
fraction of consumed atoms $R_a$ over the complete cycle.

\begin{table}[t]
    \caption{
        Finite-time cascade performance as a function of the hot
        phaseonium phase. Here
        $R_P=P_B/(2P_1)$ is the fraction of the power of two independent
        single-cavity engines retained by the cascade, while
        $R_a=M/(2N)$ is the corresponding fraction of consumed hot
        phaseonium atoms.
    }
    \label{tab:cascade-scan}
    \centering
    \small
    \setlength{\tabcolsep}{3.5pt}
    \begin{ruledtabular}
        \begin{tabular}{c rr cccc}
            $\phi_h/\pi$
             & $N$
             & $M$
             & $P_B/P_1$
             & $R_P$
             & $R_a$
             & $R_P/R_a$
            \\
            \hline
            $10/13$
             & $173\,200$
             & $237\,270$
             & $1.4763$
             & $0.7382$
             & $0.6774$
             & $1.0897$
            \\
            $2/3$
             & $87\,330$
             & $119\,640$
             & $1.4882$
             & $0.7441$
             & $0.6720$
             & $1.1072$
            \\
            $1/2$
             & $43\,880$
             & $60\,110$
             & $1.5046$
             & $0.7523$
             & $0.6647$
             & $1.1317$
            \\
            $1/3$
             & $29\,400$
             & $40\,270$
             & $1.5153$
             & $0.7576$
             & $0.6600$
             & $1.1479$
            \\
        \end{tabular}
    \end{ruledtabular}
\end{table}

The scan exhibits two consistent trends. Moving toward smaller $\phi_h$ increases the relaxation rate and reduces both cycle durations, thereby increasing the absolute powers $P_A$ and $P_B$.
Moreover, the ratio $R_P/R_a$ increases across the investigated points: reusing the atomic stream in the cascade retains a larger fraction of the two-engine power than the fraction of hot atoms it consumes.
The demonstrated advantage is therefore an improvement of power per injected hot atom for the explicitly simulated two-cavity cascade extension.

\begin{figure}[t!]
    \centering
    \includegraphics[width=.85\linewidth]{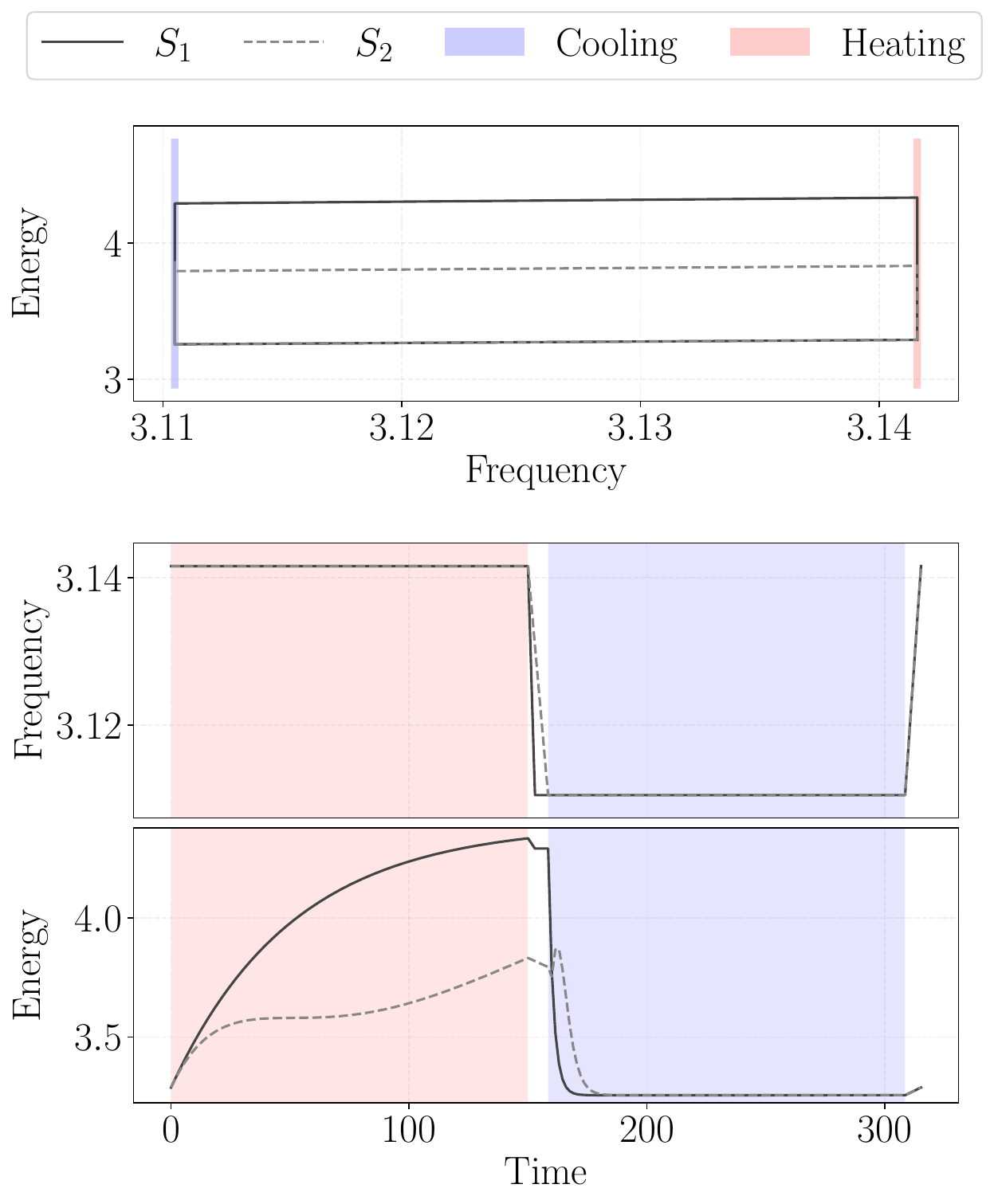}
    \caption{Finite-time dynamics of two cavities $S_1$ and $S_2$ during one thermodynamic cycle. The isochoric stroke is not sufficiently long for the second cavity to reach the stationary state, and the downstream cavity consequently performs a smaller cycle, represented by the dashed gray line. The hot and cold coherence phases are $\phi_H=0.84\pi$ and $\phi_C=\pi/40$, respectively. The mutual information during expansion is approximately $0.008$. The upper panel shows the energy--frequency cycles, while the lower panels show the frequency and energy dynamics.}
    \label{fig:correlated-cascade-cycle}
\end{figure}

The finite-time operation also generates correlations between the cavities.
If the isochoric strokes are not long enough for both cavities to reach the stationary state, the two working media perform different cycles, as shown in Fig.~\ref{fig:correlated-cascade-cycle}.
In this scenario, before the two cavities are both thermalized, they remain correlated through the action of common phaseonium atoms interacting first with the first cavity and then with the second.
This action creates a unidirectional flow of information, and thus the mutual information~\cite{zurek_information_1983} of the bipartite system will be non-zero throughout the cycle.

\begin{figure}[t!]
    \centering
    \includegraphics[width=0.63\linewidth]{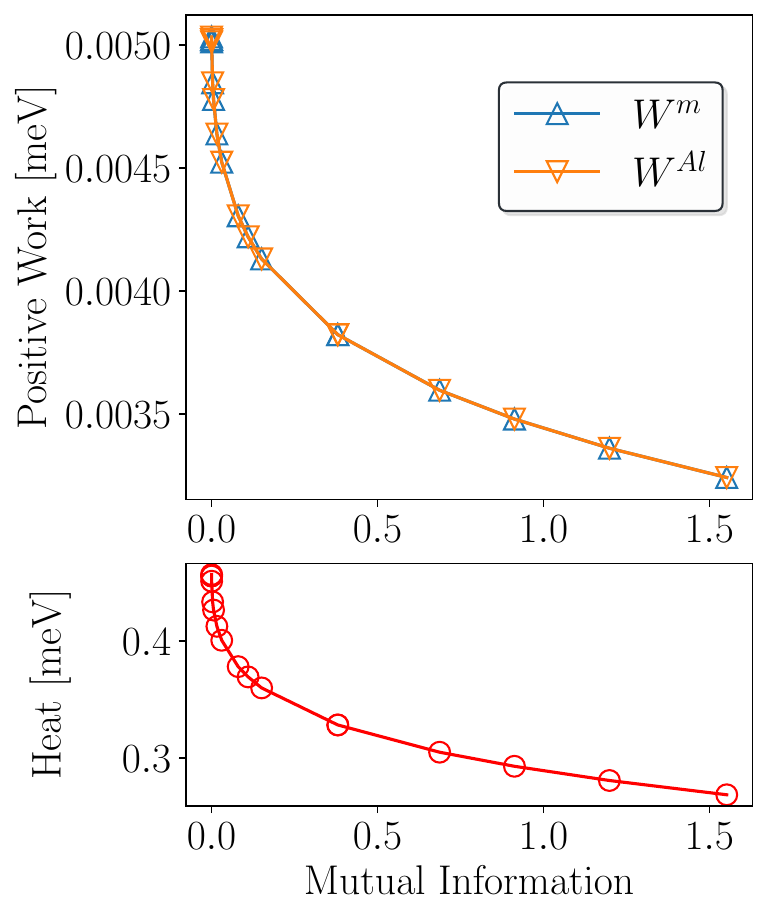}
    \caption{Mechanical work $W^m$, Alicki work $W^{\rm Al}$, and heat absorbed during the hot isochore as functions of the mutual information shared by the two cavities. Increasing mutual information in this scan corresponds to shorter thermalization times, and therefore to less absorbed heat and extracted work. The hot and cold coherence phases are $0.681\pi$ and $1.525\pi$, respectively.}
    \label{fig:mutualinfo-and-work}
\end{figure}

Figure~\ref{fig:mutualinfo-and-work} shows the work and heat exchanged as functions of the mutual information generated during partial thermalization. In this scan, larger mutual information corresponds to shorter isochoric strokes. Less heat is consequently absorbed and less work is extracted. The agreement between the mechanical work of Eq.~\eqref{def:mechanical-work} and the Alicki work of Eq.~\eqref{def:alicki-definitions} confirms that the pressure-based work accounting is consistent with the quantum first-law in the regime considered.

Since the analytical collision map employed here is derived explicitly for two cavities, we do not provide an asymptotic scaling law for larger arrays.
Nevertheless, the two-cavity cascade establishes a clear operational advantage: reusing the same atomic stream increases the absolute finite-time power and improves the power delivered per consumed phaseonium atom.

\subsection{Discussion}
The phaseonium beam is a non-thermal resource, and its apparent temperature is not the temperature of an equilibrium reservoir. It is the temperature of the Gibbs state selected by the cavity collision map, or equivalently the parameter that rewrites the gain--loss ratio in detailed-balance form.
This allows the phaseonium beam to operate as an enineered non-thermal reservoir for the cavity cycle.

The passive-energy analysis addresses this point at the level of the working fluid.
The phaseonium atoms do not deposit ergotropy in the cavity field.
Starting from diagonal thermal states, the cavity remains diagonal during the isochoric strokes, and its ergotropy is zero up to numerical precision.
Thus, the piston extracts work during the adiabatic strokes from a thermal state of the cavity, with a continuous and slow change in the eigenvalues of the Hamiltonian.

The cost analysis addresses the same issue at the level of the reservoir.
A thermal same-$T_\phi$ reference reservoir is used as benchmark for the bare Otto cycle and gives the same efficiency.
The same efficiency can be obtained with an incoherent same-population reference reservoir, used as benchmark for isolating the role of coherence in the atomic preparation.
The phaseonium route is therefore useful when the desired apparent temperatures are more naturally obtained by adding coherence to a given population distribution than by changing the populations alone.
The lower-bound analysis in Fig.~\ref{fig:reservoir-cost-difference} identifies a region in which the phaseonium route has the smaller collision-weighted preparation cost bound, when the preparation temperature $T_0$ is some order of magnitude lower than the target temperature.

Phaseonium provides a simple and tunable route to engineer apparent temperatures.
The cascade analysis extends this resource-level comparison to finite-time operation.
The downstream cavity increases both the extracted work and the number of collisions required per cycle.
Nevertheless, the additional thermalization time does not cancel the performance gain: the two-cavity engine produces approximately $47\%$--$52\%$ more power than the single-cavity engine over the investigated phase range.
At the same time, it uses only $65\%$--$68\%$ of the total atoms required by two independent engines.
The resulting $9\%$--$15\%$ enhancement of power per injected hot atom shows that the atomic stream is reused more efficiently in the cascade.
This conclusion concerns the explicitly simulated one-to-two-cavity extension and should not be interpreted as an asymptotic linear-scaling result.

In our numerical simulations, we adopt natural units, where $\hbar = k_B = 1$. In this framework, all physical quantities are scaled by the cavity's fundamental frequency $\omega_0$.
Time is expressed in units of inverse frequency ($\omega_0^{-1}$), and energy and temperature are expressed in units of frequency ($\omega_0$). The cavity length $L(t)$ is strictly defined by the fundamental resonance condition, $L(t) = 2\pi c/\omega(t)$.
This allows us to simulate the cycle dynamics in a scale-invariant manner, where results can be mapped to any physical frequency range of interest.

In our simulations, the piston displacement was kept at the percent level. Larger excursions can be obtained by changing the operating parameters, but the trade-off is time: phases that strongly modify the apparent temperature may also slow down thermalization.
Moreover, slower adiabatic motion can reduce non-adiabatic and dissipative corrections that are neglected in the present pressure model.
More refined treatments of radiation pressure and moving boundaries can be found in Refs.~\cite{machado_quantum_radiation_2002, cavalcanti_quantum_forces_2022, butera_corrections_2025}.

To provide a concrete experimental context for our results, we define a reference frequency scale $\omega_{\text{ref}}$ based on the physical dimensions of the system.
To probe our results on a macroscopic scale, our proposal is the use of macroscopic Fabry--P\'erot resonators, which currently achieve cavity lengths $L$ on the scale of millimeters to centimeters.
Accordingly, we select a reference frequency of $\omega_{\text{ref}} = \omega_0 = 2\pi \times 50$ GHz, corresponding to a fundamental mode wavelength of $\lambda \approx 6$ mm, to be consistent with the seminal work by S. Kuhr~\cite{kuhr_ultrahigh_2007}.
This choice fixes the physical scales of our dimensionless units: the unit of time maps to $t_0 = \omega_{\text{ref}}^{-1} \approx 3.18$ ps, while the unit of temperature maps to $T_0 = \hbar \omega_{\text{ref}} / k_B \approx 2.4$ K.
Consequently, a simulation temperature of $T \approx 1$ corresponds to a physical temperature of $2.4$ K, placing our engine in the standard cryogenic regime of microwave cavity QED experiments.

We also neglected photon losses by assuming ideal high-$Q$ cavities.
This is the natural first step for isolating the thermodynamic role of phaseonium coherence.
Losses and dephasing would reduce the contrast between the phaseonium and reference cycles and should be included in a device-level optimization.

Many different optomechanical platforms can be selected to perform this cycle.
Several macroscopic cavity architectures offer distinct geometric and frequency advantages for physically realizing the engine, from cantilevers to suspended mirrors to membranes~\cite{aspelmeyer_cavity_2014}.
The canonical standard is established by the superconducting Fabry--P\'erot resonators.
These can achieve cavity lengths on the scale of millimetres to centimetres, while exhibiting ultra-high quality ($Q$) factors, which indicate strong resilience to photon losses~\cite{vahala_optical_2003, fama_continuous_2024,qiao_integrated_2025}.
The canonical superconducting Fabry--P\'erot resonator~\cite{kuhr_ultrahigh_2007} operates at 51.099 GHz, utilizing an open geometry with 50 mm diameter mirrors separated by a 27.57 mm gap.
For tighter confinement, seamless 3D crossed-tube niobium cavities operate at 98.2 GHz via intersecting 1.5 mm diameter tubes, forming macroscopic cavities~\cite{suleymanzade_tunable_2020}.
To balance high cooperativity with necessary transverse optical access, modern near-confocal Fabry--P\'erot resonators target the 50 to 100 GHz band, featuring 48 mm diameter mirrors and extended macroscopic gaps up to 47.15 mm~\cite{zhang_optically_2025}.

\section{Conclusions}
We have analyzed an optomechanical Otto engine driven by phaseonium atoms, with a detailed treatment of passive energy, radiation-pressure work, reservoir-preparation costs, and scalability considerations, thanks to the use of a collision model to find the exact atom-cavity interaction equations.

The phaseonium beam sets the detailed balance of the cavity collision map through its populations and ground-state coherence. The cavity then relaxes to a Gibbs state at the corresponding apparent temperature. During this process, the field remains diagonal and passive: its ergotropy and energy-basis coherence vanish up to numerical precision, while the collision map drives it toward its Gibbs fixed point.
The engine therefore converts passive cavity energy into mechanical work through the controlled change of the cavity Hamiltonian.

The Otto cycle performed using phaseonium reservoirs reproduces the same cycle of a thermal engine operating at the same apparent hot and cold temperatures.
On the other hand, an incoherent same-population atomic beam cannot, because removing coherence changes the gain--loss balance.
Including lower bounds on preparation costs shows that phaseonium is best understood as a reservoir-engineering strategy with a population--coherence trade-off.

We have also quantified the finite-time performance of a two-cavity cascade.
Although the downstream cavity lengthens the minimum collision number required to thermalize it, this number of collisions is controlled by the absolute relaxation gap $\kappa=\Gamma_\downarrow-\Gamma_\uparrow$, again dependent on the phaseonium phase $\phi$.
After this increase in duration is included, the cascade still produces more power than the single-cavity engine. With respect to two independent single-cavity engines, the cascade retains approximately $75\%$ of the power of two independent engines while consuming only approximately $66\%$ of their total hot and cold phaseonium atoms, corresponding to an enhancement of power per injected atom over a complete cycle.

Phaseonium remains attractive because it is a minimal coherent reservoir whose effect on detailed balance is analytically transparent. This makes the proposed optomechanical engine a useful testbed for studying how microscopic coherent resources are converted into macroscopic mechanical work once thermodynamic accounting is made explicit.
\\

\begin{acknowledgments}
    R.L.F. acknowledges support by MUR (Ministero dell’Università e della Ricerca) through the following projects: PNRR Project ICON-Q – Partenariato Esteso NQSTI – PE00000023 – Spoke 2 – CUP: J13C22000680006, PNRR Project QUANTIP – Partenariato Esteso NQSTI – PE00000023 – Spoke 9 – CUP: E63C22002180006. G.A. acknowledges funding from the UK Engineering and Physical Sciences Research Council (EPSRC) under Grant No.~EP/X010929/1.
    F.A. acknowledges the kind hospitality of the University of Nottingham, where part of this work was carried out, and the valuable discussions with Tommaso Tufarelli.
\end{acknowledgments}

\bibliography{main}

\appendix

\section{Microscopic phaseonium--cavity model}
\label{app:microscopic-model}

In this Appendix, we summarize the microscopic collision model underlying the isochoric strokes.
The derivation follows Ref.~\cite{amato_heating_2024}.

The working medium is a single-mode cavity field described by the Hamiltonian
\begin{equation}
    H_S
    =
    \hbar\omega_c
    \left(
    \hat a^\dagger \hat a+\frac{1}{2}
    \right),
    \label{eq:app-cavity-hamiltonian}
\end{equation}
where $\hat a^\dagger$ and $\hat a$ are the photon creation and annihilation operators and $\omega_c$ is the cavity frequency.
For the two-cavity cascade, the free Hamiltonian is
\begin{equation}
    H_{S_1S_2} = H_{S_1}\otimes \mathbb{I}_{S_2} + \mathbb{I}_{S_1}\otimes H_{S_2},
    \label{eq:app-two-cavity-hamiltonian}
\end{equation}
with $H_{S_j} \equiv H_S$ from Eq.~\eqref{eq:app-cavity-hamiltonian} for $j=1,2$.

Each atomic ancilla is a three-level system in $\Lambda$ configuration, with excited state $\ket e$ and two degenerate ground states $\ket{g_1}$ and $\ket{g_2}$.
We introduce the transition operators
\begin{equation}
    \hat\sigma_i^+
    =
    \ket e\!\bra{g_i},
    \qquad
    \hat\sigma_i^-
    =
    \ket{g_i}\!\bra e,
    \qquad i=1,2.
    \label{eq:app-atomic-ladder-operators}
\end{equation}
Taking the energy of the ground-state doublet as zero, the atomic Hamiltonian is
\begin{equation}
    H_A
    =
    \hbar\omega_A\ket e\!\bra e
    =
    \frac{\hbar\omega_A}{2}
    \left(
    \hat\sigma_1^+\hat\sigma_1^-
    +
    \hat\sigma_2^+\hat\sigma_2^-
    \right).
    \label{eq:app-atomic-hamiltonian}
\end{equation}

All incoming atoms are identically prepared in the phaseonium state
\begin{equation}
    \lambda_\phi
    =
    \begin{pmatrix}
        \alpha^2 & 0                              & 0 \\[2pt]
        0        & \dfrac{\beta^2}{2}
                 & -i\dfrac{\beta^2}{2}e^{-i\phi}     \\[6pt]
        0        & i\dfrac{\beta^2}{2}e^{i\phi}
                 & \dfrac{\beta^2}{2}
    \end{pmatrix},
    \label{eq:app-phaseonium-state}
\end{equation}
where
\begin{equation}
    \alpha^2+\beta^2=1.
    \label{eq:app-phaseonium-normalization}
\end{equation}
The populations are therefore $\alpha^2$ in the excited state and $\beta^2/2$ in each ground state, while the parameter $\phi$ controls the phase of the coherence inside the degenerate ground-state subspace.
Complete dephasing in the atomic energy basis gives the same-population incoherent reference
\begin{equation}
    \mathcal{D}[\lambda_\phi]
    =
    \begin{pmatrix}
        \alpha^2 & 0                               & 0                               \\[2pt]
        0        & \dfrac{\beta^2}{2}              & \hspace{0.6em}0\hspace{0.6em}{} \\[6pt]
        0        & \hspace{0.6em}0\hspace{0.6em}{} & \dfrac{\beta^2}{2}
    \end{pmatrix}.
    \label{eq:app-dephased-phaseonium-state}
\end{equation}

The phase convention in Eq.~\eqref{eq:app-phaseonium-state} differs from that used in Ref.~\cite{amato_heating_2024} only by a constant shift of $\pi/2$.

Within the rotating-wave approximation, the interaction between the cavity field and one atom is
\begin{equation}
    V_{SA}
    =
    \hbar\Omega
    \left[
        \hat a
        \left(
        \hat\sigma_1^+
        +
        \hat\sigma_2^+
        \right)
        +
        \hat a^\dagger
        \left(
        \hat\sigma_1^-
        +
        \hat\sigma_2^-
        \right)
        \right],
    \label{eq:app-interaction-hamiltonian}
\end{equation}
where $\Omega$ is the atom--cavity coupling strength.
The total Hamiltonian during one collision is consequently
\begin{equation}
    H_{\mathrm{tot}}
    =
    H_S+H_A+V_{SA}.
    \label{eq:app-total-hamiltonian}
\end{equation}

We assume resonant atom--cavity interactions,
\begin{equation}
    \omega_c=\omega_A\equiv\omega.
    \label{eq:app-resonance-condition}
\end{equation}
Moving to the interaction picture with respect to $H_S+H_A$ then removes the free evolution during each collision, leaving Eq.~\eqref{eq:app-interaction-hamiltonian} as the generator of the atom--cavity dynamics.
For the cascade, the interaction with cavity $S_j$ is obtained by replacing $\hat a$ with $\hat a_j$ in Eq.~\eqref{eq:app-interaction-hamiltonian}.

\section{Single-cavity collision dynamics}
\label{app:single-cavity-map}

We now derive the reduced dynamics generated by the repeated interaction of the cavity with the phaseonium atoms.
Each incoming atom is initially uncorrelated with the cavity and interacts with it only once.
Immediately before the $k$th collision, the joint state is therefore
\begin{equation}
    \rho_S^{(k-1)}\otimes\uplambda_\phi,
    \label{eq:app-precollision-state}
\end{equation}
where $\uplambda_\phi$ is given in
Eq.~\eqref{eq:app-phaseonium-state}.

Let $\Delta t_{\mathrm{int}}$ be the duration of a single atom--cavity interaction and define the accumulated Rabi phase
\begin{equation}
    \theta=\Omega\Delta t_{\mathrm{int}}.
    \label{eq:app-rabi-phase}
\end{equation}
In the interaction picture, the collision unitary is
\begin{equation}
    U_{SA}
    =
    \exp\left(
    -\frac{i}{\hbar}V_{SA}\Delta t_{\mathrm{int}}
    \right).
    \label{eq:app-collision-unitary}
\end{equation}
Introducing the dimensionless interaction operator $\widetilde V_{SA}=V_{SA}/(\hbar\Omega)$, this can equivalently be written as $U_{SA}=\exp(-i\theta\widetilde V_{SA})$.

In the atomic basis $\{\ket e,\ket{g_1},\ket{g_2}\}$, the exact unitary for an arbitrary interaction time takes the form
\begin{equation}
    U_{SA}
    =
    \begin{pmatrix}
        \hat C
         &
        -i\hat S^\dagger
         &
        -i\hat S^\dagger
        \\[3pt]
        -i\hat S
         &
        \dfrac{\hat C'+\mathbb{I}}{2}
         &
        \dfrac{\hat C'-\mathbb{I}}{2}
        \\[7pt]
        -i\hat S
         &
        \dfrac{\hat C'-\mathbb{I}}{2}
         &
        \dfrac{\hat C'+\mathbb{I}}{2}
    \end{pmatrix},
    \label{eq:app-exact-collision-unitary}
\end{equation}
where the entries are operators acting on the cavity Hilbert space,
\begin{align}
    \hat C
     & =
    \cos\left(
    \theta\sqrt{2\hat a\hat a^\dagger}
    \right),
    \label{eq:app-C-operator}
    \\
    \hat C'
     & =
    \cos\left(
    \theta\sqrt{2\hat a^\dagger\hat a}
    \right),
    \label{eq:app-Cprime-operator}
    \\
    \hat S
     & =
    \hat a^\dagger
    \frac{
        \sin\left(
        \theta\sqrt{2\hat a\hat a^\dagger}
        \right)
    }{
        \sqrt{2\hat a\hat a^\dagger}
    }.
    \label{eq:app-S-operator}
\end{align}
These operators satisfy
\begin{equation}
    \hat C^2+2\hat S^\dagger\hat S=\mathbb{I},
    \qquad
    \hat C'^2+2\hat S\hat S^\dagger=\mathbb{I}.
    \label{eq:app-photonic-identities}
\end{equation}

After the collision, the atomic degrees of freedom are traced out.
The reduced cavity state is consequently updated through the collision map
\begin{equation}
    \rho_S^{(k)}
    =
    \mathcal{E}_\phi
    \left[
        \rho_S^{(k-1)}
        \right]
    =
    \Tr_A
    \left[
        U_{SA}
        \left(
        \rho_S^{(k-1)}\otimes\lambda_\phi
        \right)
        U_{SA}^\dagger
        \right].
    \label{eq:app-single-cavity-map}
\end{equation}
The map admits the Kraus representation
\begin{equation}
    \mathcal{E}_\phi[\rho_S]
    =
    \sum_{\mu=0}^{4}
    \hat E_\mu\rho_S\hat E_\mu^\dagger,
    \label{eq:app-kraus-map}
\end{equation}
with
\begin{align}
    \hat E_0
     & =
    \sqrt{
        1-\frac{\gamma_\alpha}{2}
        -\frac{\gamma_\beta}{2}
    }\,
    \mathbb{I},
    \\
    \hat E_1
     & =
    \sqrt{\frac{\gamma_\alpha}{2}}\,
    \hat C,
    \\
    \hat E_2
     & =
    \sqrt{\gamma_\alpha}\,
    \hat S,
    \\
    \hat E_3
     & =
    \sqrt{\frac{\gamma_\beta}{2}}\,
    \hat C',
    \\
    \hat E_4
     & =
    \sqrt{\gamma_\beta}\,
    \hat S^\dagger.
\end{align}
The two coefficients appearing in the map are the relevant atomic second moments,
\begin{align}
    \gamma_\alpha
     & =
    \Tr_A
    \left[
        \left(
        \hat\sigma_1^+
        +
        \hat\sigma_2^+
        \right)
        \left(
        \hat\sigma_1^-
        +
        \hat\sigma_2^-
        \right)
        \uplambda_\phi
        \right]
    =
    2\alpha^2,
    \label{eq:app-gamma-alpha}
    \\
    \gamma_\beta
     & =
    \Tr_A
    \left[
        \left(
        \hat\sigma_1^-
        +
        \hat\sigma_2^-
        \right)
        \left(
        \hat\sigma_1^+
        +
        \hat\sigma_2^+
        \right)
        \uplambda_\phi
        \right]
    =
    \beta^2(1-\sin\phi).
    \label{eq:app-gamma-beta}
\end{align}
The phase dependence of the dynamics is therefore entirely contained in $\gamma_\beta$. In particular, the coherence between the degenerate ground states modifies the atomic absorption channel, while $\gamma_\alpha$ depends only on the excited-state population.

Using Eq.~\eqref{eq:app-photonic-identities}, the exact finite-difference equation can also be written as
\begin{equation}
    \rho_S^{(k)}-\rho_S^{(k-1)}
    =
    \sum_{\mu=1}^{4}
    \mathcal{D}[\hat E_\mu]
    \rho_S^{(k-1)},
    \label{eq:app-finite-difference-master-equation}
\end{equation}
where
\begin{equation}
    \mathcal{D}[\hat L]\rho
    =
    \hat L\rho\hat L^\dagger
    -
    \frac{1}{2}
    \left\{
    \hat L^\dagger\hat L,\rho
    \right\}
    \label{eq:app-dissipator}
\end{equation}
is the standard dissipator.

The structure of the Kraus operators also shows that the map preserves diagonality in the cavity number basis.
The operators $\hat C$ and $\hat C'$ are diagonal functions of the photon-number operator, while $\hat S$ and $\hat S^\dagger$ change the photon number by one.
Consequently, if the cavity enters a collision in a state diagonal in the energy basis, the state remains diagonal after the collision.
This property underlies the absence of cavity energy-basis coherence discussed in Sec.~\ref{sec:thermodynamic-interpretation}.

For weak individual collisions, Eq.~\eqref{eq:app-single-cavity-map} reduces to the usual continuous-time master equation.
Expanding the operators in Eq.~\eqref{eq:app-C-operator}, \eqref{eq:app-Cprime-operator}, and~\eqref{eq:app-S-operator} to second order in $\theta$ gives
\begin{equation}
    \frac{d\rho_S}{dt}
    =
    \Gamma_\downarrow
    \mathcal{D}[\hat a]\rho_S
    +
    \Gamma_\uparrow
    \mathcal{D}[\hat a^\dagger]\rho_S,
    \label{eq:app-continuous-master-equation}
\end{equation}
where
\begin{equation}
    \Gamma_\uparrow
    =
    \Gamma_0\gamma_\alpha,
    \qquad
    \Gamma_\downarrow
    =
    \Gamma_0\gamma_\beta.
    \label{eq:app-excitation-deexcitation-rates}
\end{equation}
Here $\Gamma_0$ is a common positive rate determined by the
atom--cavity coupling, the interaction time, and the atomic injection
rate. For an injection rate $\nu$, the weak-collision expansion gives
$\Gamma_0=\nu\theta^2$. Since this prefactor is common to the two
channels, it does not affect their detailed-balance ratio. In
collision units, the same dynamics can be characterized directly by
$\Gamma_\uparrow=\gamma_\alpha$ and
$\Gamma_\downarrow=\gamma_\beta$.

The stationary state $\rho_S^*$ is the fixed point of the collision
map,
\begin{equation}
    \mathcal{E}_\phi[\rho_S^*]=\rho_S^*.
    \label{eq:app-fixed-point-condition}
\end{equation}
Writing it as
\begin{equation}
    \rho_S^*
    =
    \sum_{n=0}^{\infty}
    p_n^*\ket n\!\bra n,
    \label{eq:app-diagonal-fixed-point}
\end{equation}
the stationary condition gives the recurrence relation
\begin{equation}
    \frac{p_{n+1}^*}{p_n^*}
    =
    \frac{\gamma_\alpha}{\gamma_\beta}
    =
    \frac{\Gamma_\uparrow}{\Gamma_\downarrow}.
    \label{eq:app-stationary-recurrence}
\end{equation}
For $\gamma_\alpha<\gamma_\beta$, the normalizable solution is
\begin{equation}
    \rho_S^*
    =
    \frac{1}{Z_\phi}
    \sum_{n=0}^{\infty}
    \left(
    \frac{\gamma_\alpha}{\gamma_\beta}
    \right)^n
    \ket n\!\bra n,
    \label{eq:app-cavity-fixed-point}
\end{equation}
with
\begin{equation}
    Z_\phi
    =
    \left(
    1-\frac{\gamma_\alpha}{\gamma_\beta}
    \right)^{-1}.
    \label{eq:app-fixed-point-normalization}
\end{equation}
Comparison with the Gibbs distribution gives
\begin{equation}
    \frac{\gamma_\alpha}{\gamma_\beta}
    =
    \exp\left(
    -\frac{\hbar\omega}{k_BT_\phi}
    \right),
    \label{eq:app-fixed-point-detailed-balance}
\end{equation}
and therefore
\begin{equation}
    k_BT_\phi
    =
    \frac{\hbar\omega}{
        \ln\left[
            \dfrac{\beta^2(1-\sin\phi)}
            {2\alpha^2}
            \right]
    }.
    \label{eq:app-derived-apparent-temperature}
\end{equation}
The temperature $T_\phi$ is thus not the temperature of the incoming
atomic state. It is the temperature of the Gibbs fixed point selected
by the detailed balance of the reduced cavity dynamics.

For the dephased same-population preparation,
Eq.~\eqref{eq:app-dephased-phaseonium-state}, the coherence
contribution vanishes and
\begin{equation}
    \gamma_\beta=\beta^2.
    \label{eq:app-classical-gamma-beta}
\end{equation}
The corresponding stationary temperature is
\begin{equation}
    k_BT_{cl}
    =
    \frac{\hbar\omega}{
        \ln\left(
        \dfrac{\beta^2}{2\alpha^2}
        \right)
    },
    \label{eq:app-derived-classical-temperature}
\end{equation}
recovering Eq.~\eqref{def:classic-temperature}.

While the ratio of the two rates determines the fixed-point
temperature, their difference determines the relaxation scale. We
define the relaxation gap as
\begin{equation}
    \kappa
    =
    \Gamma_\downarrow-\Gamma_\uparrow
    =
    \Gamma_0
    \left[
        \beta^2(1-\sin\phi)-2\alpha^2
        \right].
    \label{eq:app-relaxation-gap}
\end{equation}
Indeed, Eq.~\eqref{eq:app-continuous-master-equation} gives
\begin{equation}
    \frac{d}{dt}
    \langle\hat n\rangle
    =
    -\kappa\langle\hat n\rangle
    +
    \Gamma_\uparrow,
    \label{eq:app-photon-number-relaxation}
\end{equation}
so that
\begin{equation}
    \langle\hat n\rangle^*
    =
    \frac{\Gamma_\uparrow}{
        \Gamma_\downarrow-\Gamma_\uparrow
    }.
    \label{eq:app-stationary-photon-number}
\end{equation}
Therefore, two atomic preparations may generate the same apparent temperature, and hence the same stationary state, while producing different relaxation gaps and different finite-time collision numbers.

\section{Two-cavity cascade collision dynamics}
\label{app:cascade-map}

We now consider two identical single-mode cavities, $S_1$ and $S_2$, arranged in cascade and driven by the same stream of phaseonium atoms.
Each incoming atom interacts first with $S_1$ and then with $S_2$, as shown in Fig.~\ref{sfig:cascade-setup}.
The complete finite-time collision map follows the derivation of Ref.~\cite{amato_heating_2024}.

During the isochoric strokes, the two cavities have the same frequency and interact with each atom with the same coupling strength and interaction time.
The local interaction Hamiltonians are
\begin{equation}
    V_{S_jA}
    =
    \hbar\Omega
    \left[
        \hat a_j
        \left(
        \hat\sigma_1^+
        +
        \hat\sigma_2^+
        \right)
        +
        \hat a_j^\dagger
        \left(
        \hat\sigma_1^-
        +
        \hat\sigma_2^-
        \right)
        \right],
    \quad j=1,2,
    \label{eq:app-cascade-interaction-hamiltonians}
\end{equation}
and the corresponding collision unitaries are
\begin{equation}
    U_{jA}
    =
    \exp\left(
    -\frac{i}{\hbar}
    V_{S_jA}\Delta t_{\mathrm{int}}
    \right).
    \label{eq:app-cascade-local-unitaries}
\end{equation}
The unitary associated with one complete cascade collision is therefore
\begin{equation}
    U_{\mathrm{cas}}
    =
    U_{2A}U_{1A}.
    \label{eq:app-cascade-unitary}
\end{equation}
The ordering in Eq.~\eqref{eq:app-cascade-unitary} fixes the direction of the cascade: the atom interacts with the upstream cavity before reaching the downstream one.

Let $\rho_{12}^{(k-1)}$ denote the joint state of the two cavities immediately before the $k$th atom enters the cascade.
Since each incoming atom is initially uncorrelated with the cavities, the global precollision state is
\begin{equation}
    \rho_{12}^{(k-1)}\otimes\uplambda_\phi.
    \label{eq:app-cascade-precollision-state}
\end{equation}
After the two sequential interactions, the atomic degrees of freedom
are discarded and the cavity state becomes
\begin{align}
    \rho_{12}^{(k)}
     & =
    \mathcal{E}_{12,\phi}
    \left[
        \rho_{12}^{(k-1)}
        \right]
    \nonumber \\
     & =
    \Tr_A
    \left[
        U_{2A}U_{1A}
        \left(
        \rho_{12}^{(k-1)}
        \otimes\uplambda_\phi
        \right)
        U_{1A}^\dagger U_{2A}^\dagger
        \right].
    \label{eq:app-cascade-collision-map}
\end{align}
Equation~\eqref{eq:app-cascade-collision-map} defines a completely positive and trace-preserving map on the joint two-cavity system.

To write the map explicitly, we introduce the local operators
\begin{subequations}
    \label{eq:app-cascade-photonic-operators}
    \begin{align}
        \hat C_j
         & =
        \cos\left(
        \theta\sqrt{2\hat a_j\hat a_j^\dagger}
        \right),
        \\
        \hat C'_j
         & =
        \cos\left(
        \theta\sqrt{2\hat a_j^\dagger\hat a_j}
        \right),
        \\
        \hat S_j
         & =
        \hat a_j^\dagger
        \frac{
            \sin\left(
            \theta\sqrt{2\hat a_j\hat a_j^\dagger}
            \right)
        }{
            \sqrt{2\hat a_j\hat a_j^\dagger}
        },
        \qquad j=1,2,
    \end{align}
\end{subequations}
where $\theta=\Omega\Delta_{\mathrm{int}}$, as in Eq.~\eqref{eq:app-rabi-phase}. The tensor products below are ordered as $\mathcal{H}_{S_1}\otimes\mathcal{H}_{S_2}$.

The cascade map admits the Kraus representation
\begin{equation}
    \mathcal{E}_{12,\phi}[\rho_{12}]
    =
    \sum_{\mu=0}^{4}
    \hat F_\mu
    \rho_{12}
    \hat F_\mu^\dagger,
    \label{eq:app-cascade-kraus-map}
\end{equation}
where
\begin{subequations}
    \label{eq:app-cascade-kraus-operators}
    \begin{align}
        \hat F_0
         & =
        \sqrt{
            1-\frac{\gamma_\alpha}{2}
            -\frac{\gamma_\beta}{2}
        }\,
        \mathbb{I}_{12},
        \label{eq:app-cascade-F0}
        \\
        \hat F_1
         & =
        \sqrt{\frac{\gamma_\alpha}{2}}
        \left(
        \hat C_1\otimes\hat C_2
        -
        2\hat S_1\otimes\hat S_2^\dagger
        \right),
        \label{eq:app-cascade-F1}
        \\
        \hat F_2
         & =
        \sqrt{\gamma_\alpha}
        \left(
        \hat S_1\otimes\hat C'_2
        +
        \hat C_1\otimes\hat S_2
        \right),
        \label{eq:app-cascade-F2}
        \\
        \hat F_3
         & =
        \sqrt{\gamma_\beta}
        \left(
        \hat S_1^\dagger\otimes\hat C_2
        +
        \hat C'_1\otimes\hat S_2^\dagger
        \right),
        \label{eq:app-cascade-F3}
        \\
        \hat F_4
         & =
        \sqrt{\frac{\gamma_\beta}{2}}
        \left(
        \hat C'_1\otimes\hat C'_2
        -
        2\hat S_1^\dagger\otimes\hat S_2
        \right).
        \label{eq:app-cascade-F4}
    \end{align}
\end{subequations}
Here
\begin{equation}
    \mathbb{I}_{12}=\mathbb{I}_{S_1}\otimes \mathbb{I}_{S_2},
\end{equation}
and the coefficients are the same as in the single-cavity map,
\begin{equation}
    \gamma_\alpha=2\alpha^2,
    \qquad
    \gamma_\beta=\beta^2(1-\sin\phi).
    \label{eq:app-cascade-gamma-coefficients}
\end{equation}
The Kraus operators satisfy
\begin{equation}
    \sum_{\mu=0}^{4}
    \hat F_\mu^\dagger\hat F_\mu
    =
    \mathbb{I}_{12},
    \label{eq:app-cascade-completeness}
\end{equation}
ensuring trace preservation.

The corresponding exact finite-difference equation is
\begin{equation}
    \rho_{12}^{(k)}
    -
    \rho_{12}^{(k-1)}
    =
    \sum_{\mu=1}^{4}
    \mathcal{D}[\hat F_\mu]
    \rho_{12}^{(k-1)},
    \label{eq:app-cascade-finite-difference}
\end{equation}
with the dissipator defined in Eq.~\eqref{eq:app-dissipator}.
Unlike the single-cavity map, the operators in Eq.~\eqref{eq:app-cascade-kraus-operators} contain nonlocal products of cavity operators.
The common atoms can therefore generate correlations between $S_1$ and $S_2$ during the isochoric evolution.

The unidirectionality of the dynamics becomes explicit by considering the two reduced cavity states
\begin{equation}
    \rho_j^{(k)}
    =
    \Tr_{S_{\bar j}}
    \left[
        \rho_{12}^{(k)}
        \right],
    \qquad j=1,2.
    \label{eq:app-cascade-reduced-states}
\end{equation}
For the upstream cavity, tracing
Eq.~\eqref{eq:app-cascade-collision-map} over $S_2$ gives
\begin{equation}
    \begin{split}
        \rho_1^{(k)}
         & =
        \Tr_{S_2A}
        \left[
            U_{2A}U_{1A}
            \left(
            \rho_{12}^{(k-1)}
            \otimes\uplambda_\phi
            \right)
            U_{1A}^\dagger U_{2A}^\dagger
            \right]
        \\
         & =
        \Tr_A
        \left[
            U_{1A}
            \left(
            \rho_1^{(k-1)}
            \otimes\uplambda_\phi
            \right)
            U_{1A}^\dagger
            \right]
        \\
         & =
        \mathcal{E}_\phi
        \left[
            \rho_1^{(k-1)}
            \right].
        \label{eq:app-upstream-reduced-map}
    \end{split}
\end{equation}
The interaction occurring afterwards between the atom and $S_2$ cannot affect the reduced state of $S_1$.
The first cavity therefore follows exactly the single-cavity map derived in Sec.~\ref{app:single-cavity-map}, independently of the state of the downstream cavity.

The situation is different for $S_2$. After the atom has interacted with $S_1$, the joint state entering the second interaction is
\begin{equation}
    \chi_{2A}^{(k)}
    =
    \Tr_{S_1}
    \left[
        U_{1A}
        \left(
        \rho_{12}^{(k-1)}
        \otimes\lambda_\phi
        \right)
        U_{1A}^\dagger
        \right].
    \label{eq:app-intermediate-second-cavity-state}
\end{equation}
The downstream cavity then evolves as
\begin{equation}
    \rho_2^{(k)}
    =
    \Tr_A
    \left[
        U_{2A}
        \chi_{2A}^{(k)}
        U_{2A}^\dagger
        \right].
    \label{eq:app-downstream-reduced-dynamics}
\end{equation}
The intermediate state $\chi_{2A}^{(k)}$ depends on the joint two-cavity state $\rho_{12}^{(k-1)}$, including the correlations created during the preceding collisions.
Consequently, the evolution of $S_2$ cannot in general be expressed as an autonomous completely positive map acting only on $\rho_2^{(k-1)}$.
Its reduced dynamics retains information about the upstream cavity and is therefore effectively non-Markovian.

The atomic stream thus mediates a one-way flow of information from $S_1$ to $S_2$: the dynamics of $S_1$ is independent of $S_2$, whereas the evolution of $S_2$ depends on the state and previous evolution of $S_1$.
No direct cavity--cavity interaction is present.
The correlations observed during finite-time operation arise solely from the sequential interaction with the common atoms.

In the mixing regime $\gamma_\alpha<\gamma_\beta$, the first cavity converges to the Gibbs fixed point Eq.~\eqref{eq:app-cavity-fixed-point}. The atoms leaving the first cavity continue to drive the downstream cavity toward the same local fixed point, so that
\begin{equation}
    \rho_1^{(k)}
    \longrightarrow
    \rho_S^*,
    \qquad
    \rho_2^{(k)}
    \longrightarrow
    \rho_S^*,
    \label{eq:app-cascade-local-fixed-points}
\end{equation}
where $\rho_S^*$ has the apparent temperature $T_\phi$ given by Eq.~\eqref{eq:app-derived-apparent-temperature}.
The convergence of the second cavity is generally slower because it interacts with atoms that have already been modified by the upstream collision.
When an isochoric stroke is stopped before this convergence is complete, the two cavities leave the stroke in different local states and may retain nonzero mutual information, as discussed in Sec.~\ref{sec:cascade-performance}.

\end{document}